\documentclass[%
 reprint,
superscriptaddress,
 amsmath,amssymb,
 aps,
]{revtex4-2}

\usepackage{graphicx}%
\usepackage{dcolumn}%
\usepackage{bm}%
\usepackage{orcidlink}
\usepackage{hyperref}%

\usepackage{xcolor}
\definecolor{revisioncolor}{RGB}{0,70,180}

\begin{document}

\preprint{APS/123-QED}

\title{Stochastic Ising networks with asymmetric and time-delayed interactions}%

\author{Hantao Zhang\,\orcidlink{0000-0001-6281-6708}}
\affiliation{School of Engineering and Applied Science, The George Washington University, Washington, DC 20052, USA}
\affiliation{Associate, Physical Measurement Laboratory, National Institute of Standards and Technology, Gaithersburg, Maryland 20899, USA}

\author{Sidra Gibeault\,\orcidlink{0009-0005-0103-7168}}
\affiliation{Institute for Research in Electronics and Applied Physics, University of Maryland, College Park, Maryland 20742, USA}
\affiliation{Associate, Physical Measurement Laboratory, National Institute of Standards and Technology, Gaithersburg, Maryland 20899, USA}

\author{Matthew W. Daniels\,\orcidlink{0000-0002-3390-4714}}
\affiliation{Physical Measurement Laboratory, National Institute of Standards and Technology, Gaithersburg, Maryland 20899, USA}

\author{Philippe Talatchian\,\orcidlink{0000-0003-2034-6140}}
\affiliation{Univ. Grenoble Alpes, CEA, CNRS, Grenoble INP, SPINTEC, 38000 Grenoble, France}

\author{Ursula Ebels\,\orcidlink{0000-0001-5061-5538}}
\affiliation{Univ. Grenoble Alpes, CEA, CNRS, Grenoble INP, SPINTEC, 38000 Grenoble, France}

\author{Advait Madhavan\,\orcidlink{0000-0002-4121-1336}}
\affiliation{Physical Measurement Laboratory, National Institute of Standards and Technology, Gaithersburg, Maryland 20899, USA}

\author{Mark D. Stiles\,\orcidlink{0000-0001-8238-4156}}
\affiliation{Physical Measurement Laboratory, National Institute of Standards and Technology, Gaithersburg, Maryland 20899, USA}

\date{\today}%

\begin{abstract}
Stochastic Ising networks are widely used to describe collective dynamics in complex systems and to compute, yet physical implementations of these networks often contain both asymmetric interactions and nonzero signal propagation times that fall outside conventional theories of their behavior. Here, we study stochastic Ising networks with asymmetric and time-delayed interactions motivated by experimental observations in coupled superparamagnetic tunnel junctions. Time delay fundamentally reshapes the dynamics induced by antisymmetric couplings, producing strong oscillatory temporal correlations consistent with experiment. Sufficiently long delays strongly suppress the non-uniformity of steady-state probability distributions, a behavior explained by an exponential suppression of the first-order equal-time cross-correlation between pairs of spins. Residual non-uniformity is due to the equal-time cross-correlation mediated by a third spin that contributes only at higher order in the coupling strength. These relatively featureless probability distributions coexist with pronounced delayed temporal correlations, distinguishing them from equilibrium high-temperature behavior. We further show analytically that these delay-induced near-uniform distributions emerge in a broad class of stochastic networks that lack symmetry-breaking bias fields. Introducing symmetry-breaking bias fields restores a strong dependence of the steady state on interactions with qualitatively modified behavior. Simulations of networks with five coupled spins demonstrate that these effects persist beyond minimal systems with only two spins. %

\end{abstract}

\maketitle

\section{Introduction} \label{sec:intro}

Stochastic Ising networks are powerful tools for both brain-inspired computation and the modeling of complex systems. By encoding information in probabilistic dynamics and complicated network structures, they enable efficient exploration of high dimensional state spaces and have been widely employed in optimization, machine learning, and inference~\cite{hopfield1982neural,ackley1985learning,kirkpatrick1983optimization,lucas2014ising,hertz2018introduction}. At the same time, they provide a unifying language for collective behavior across disciplines, from magnetic materials~\cite{ising1925beitrag,edwards1975theory,sherrington1975solvable} to biological~\cite{dayan2005theoretical,schneidman2006weak,meshulam2025statistical} or social networks~\cite{galam2008sociophysics,castellano2009statistical}.

Stochastic Ising networks without delay or asymmetry form the basis for Ising machines applied to solving hard optimization problems~\cite{grollier2016spintronic,grollier2020neuromorphic,mohseni2022ising,zhang2024review,lee2026fundamental} and energy-based machine learning algorithms~\cite{hopfield1982neural,ackley1985learning,carleo2019machine}. A well-studied regime for these systems corresponds to models whose statistical properties are described through well-defined energy functionals, and readily submit to analysis by the usual tools of statistical mechanics.
At the opposite extreme, asymmetric or time-delayed interactions give rise to intrinsically non-Hamiltonian dynamics; such systems lack energy functionals, and the ordinary equilibrium theory of statistical mechanics fails to correctly describe the networks. Asymmetry and delay have each been studied in isolation. For example, Ising systems whose spin interactions are asymmetric, sometimes referred to as asymmetric, kinetic, or non-reciprocal Ising models, have been investigated with respect to their phase transitions, order parameter dynamics, and other system properties~\cite{crisanti1988dynamics,mezard2011exact,coolen1996dynamical,aguilera2021unifying,guislain2024collective,avni2025nonreciprocal,avni2025dynamical,blom2025local,weiderpass2025solving,di2025off,ishihara2025state,xue2025critical}. Ising systems where multiple blocks of Ising spins with symmetric interactions are connected through directed and highly asymmetric interactions, have been studied in the context of invertible Boolean logic~\cite{camsari2017stochastic,camsari2019p,chowdhury2023full}. Prior studies of time-delayed Ising systems usually model each Ising spin as a bistable continuous variable with its own and other spins' delayed states as input, but where the couplings are symmetric and the joint states of all spins are simplified through a mean field treatment~\cite{tsimring2001noise,masoller2003distribution,huber2003dynamics,huber2005cooperative}. Other early works about time-delayed Ising systems similarly employed mean-field treatments or else modeled delayed interaction with an integration over past states, sidestepping issues of infinite dimensionality typically associated with discrete delay times~\cite{sompolinsky1986temporal,schulz2002kinetically,caccioli2008ising,touboul2014hipster,collet2016rhythmic}. 

While these works addressed important problems in specificity, they do not generally treat the stochastic Ising networks that serve as the basis for probabilistic computing. Prior time-delayed Ising works often oversimplify joint spin states or coupled spins, and overlook the fact that delay times in probabilistic hardware usually lack explicit integration kernels. More importantly, the combined effect of asymmetry and delay on probability distributions over state spaces and temporal correlations of dynamics---two central metrics for stochastic Ising network behavior---are less studied, and there is no research examining the effects of these two nonidealities simultaneously. Most biological systems on which these models are based---and implementations of such approaches in hardware---operate in the presence of both delay and asymmetry. Even implementations of Ising machines purpose-built to minimize energy functionals are subject to moderate imperfections in fabrication that inevitably yield asymmetric coupling and nonzero signal propagation times~\cite{yamamoto2017coherent,yamamoto2020coherent,aadit2022massively,gibeault2024programmable,gao2024photonic,lee2026fundamental,onizawa2026unified}. Some preliminary evidence suggests that asymmetry and time delay may address the question of broken time-reversal symmetry in biological neural networks~\cite{meshulam2025statistical}, and that properly engineered time delays may be used to implement optimization algorithms~\cite{selcuk2025dac} or accelerate neuromorphic computation~\cite{aadit2022massively}. In these regards, it is critical to develop a general Ising framework combining both phenomena, in order to reveal the relationships among asymmetric time-delayed interactions, probability distributions, and temporal correlations.

Motivated by the observation of oscillatory temporal correlations in a hardware implementation of two coupled spins with fully antisymmetric coupling and finite signal propagation time, we investigate the interplay between asymmetric interactions and time delays within a generalized Ising network. We show that this interplay gives rise to qualitatively new behavior that cannot be inferred from either ingredient alone. In particular, we find that when interactions are both asymmetric and time-delayed, the delayed spin correlations exhibit strong temporal oscillations in both experiment and theory. Furthermore, a sufficiently long time delay suppresses the first-order dependence of the spins' steady-state joint probability distribution on the interaction strength, leaving a statistical trace observable only at second-order. We explain the mechanism of suppressed non-uniformity of probability distributions with a perturbative theory: the direct equal-time cross-correlation between any two spins---which is first-order in the coupling strength and is typically responsible for non-uniformity of the distribution---exponentially decays in the limit of long interaction delays. What survives in this limit is an indirect correlation, mediated by a third spin, which manifests only at second-order in the coupling strength.

Our observed near-uniform joint probability distributions of spin states are accompanied by oscillatory temporal correlations, distinct from the randomization induced by high temperature or weak coupling, demonstrating that strong dynamical structure persists even when steady-state distributions appear featureless. We show that nearly uniform steady states arising from interaction delays is a general feature for a wide range of systems with specific symmetries, including, but not limited to, Potts-model, Kuramoto-model, and Heisenberg-model systems. We further show that breaking the spin inversion symmetry (\textit{i.e.} $\mathbb{Z}_2$ symmetry) via bias fields restores a sensitivity of the distribution to exchange interactions, albeit in a manner qualitatively different than under instantaneous interactions. In addition, long delays induce plateau-like behavior in the evolution of probability distributions, indicating the sensitivity of dynamics to initial conditions. These results are supported by general theoretical arguments and numerical demonstrations beyond minimal two-spin systems, confirming the validity of our results for general stochastic Ising networks with asymmetric and time-delayed interactions. Our findings reveal that asymmetry and delay, often regarded as imperfections, can potentially be harnessed to enable novel functionality in probabilistic computing hardware, with implications for the modeling of complex systems, though specific demonstrations of these functionalities as well as discussions of their pros and cons are left to future work.

\section{Experimental Motivation} \label{sec:motivation}
\begin{figure}[!htbp]
\centering
\includegraphics[width=0.9\linewidth]{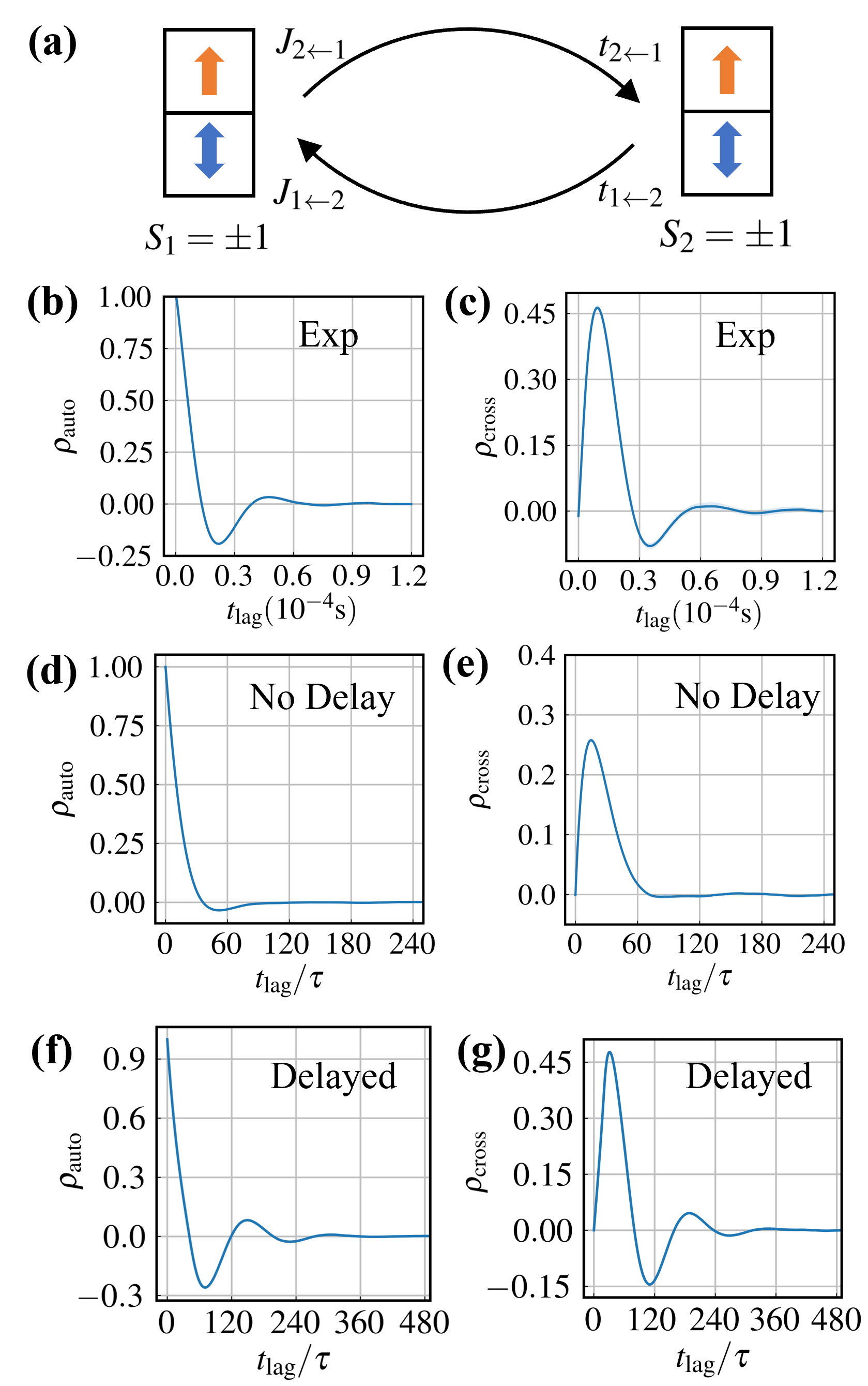}
\caption{(a) Two coupled SMTJs, modeled as two Ising spins $S_1$ and $S_2$, with interaction strengths $J_{1 \leftarrow 2}$ and $J_{2 \leftarrow 1}$, as well as finite time delays $t_{1 \leftarrow 2}$ and $t_{2 \leftarrow 1}$. Experimental auto- and cross-correlation data are shown in (b) and (c) respectively, with maximum coupling strength allowed by the circuit and $J_{1\leftarrow 2}\approx-J_{2\leftarrow 1}$. Results from simulations with antisymmetric coupling, for instantaneous interactions is shown in (d,e) and time-delayed interactions in (f,g) for the autocorrelation function (d,f) and for the cross-correlation function (e,g) as a function of $t_{\rm lag}$, the time difference between measurements of digitized SMTJ voltage states or spin states. The intrinsic characteristic time scale of each spin in the model is $\tau = 1/\lambda_{\rm flip}^{0}$. $95~\%$ confidence intervals of correlation functions are obtained from standard statistical analysis but are narrower than the width of the line. Simulations are performed with $\tau = 4.98~\Delta t$, $\Delta E_{i} = 4~k T$, $J_{1 \leftarrow 2} = - J_{2 \leftarrow 1} = k T$, $h_1 = h_2 = 0$, $t_{1 \leftarrow 2} = t_{2 \leftarrow 1} = 0$ for (d) and (e) or $t_{1 \leftarrow 2} = t_{2 \leftarrow 1} =100~\Delta t$ for (f) and (g). Each data point gives the mean correlation averaged over $10^{4}$ ensembles and for $5000~\Delta t$ after reaching steady state. Details of simulation parameters are given in Section~\ref{sec:model}.}
\label{fig:motivation_model}
\end{figure}

Our study of asymmetric and time-delayed interactions in Ising models is motivated by an experiment measuring coupled superparamagnetic tunnel junctions (SMTJs) with the setup described in Ref.~\cite{gibeault2024programmable}. The experiment is based on electrically coupling two SMTJs, as shown schematically in Fig.~\ref{fig:motivation_model}(a). Each SMTJ is modeled as an Ising spin $S_{i}$, with $S_{i} = \pm 1$ corresponding to the parallel or antiparallel configuration of magnetizations, respectively. The electric current from SMTJ $j$ to SMTJ $i$, which controls the probability distribution of magnetic configurations for SMTJ $i$, is abstracted as a unidirectional coupling strength $J_{i \leftarrow j}$. Published results in Ref.~\cite{gibeault2024programmable} consider the cases of symmetric ferromagnetic and symmetric antiferromagnetic coupling, and analyze the results with a Markov model based on treating the coupling as instantaneous. Agreement between the experiment and the model is good. 

Here, we report unpublished results from a similar measurement conducted at the time of the experiment in Ref.~\cite{gibeault2024programmable}, but where the coupling is antisymmetric: one SMTJ applies a ferromagnetically-aligning current to the other, but this latter SMTJ is applies an antiferromagnetically-aligning current back to the first SMTJ. In brief, $J_{1 \leftarrow 2} \approx -J_{2 \leftarrow 1}$. We digitize the voltage states of SMTJ $i$ as $S_i=\pm 1$ and define the autocorrelation function as the correlation between $S_{i}(t)$ and $S_{i}(t-t_{\rm lag})$, while the cross-correlation function is defined as the correlation between $S_{i}(t)$ and $S_{j\neq i}(t-t_{\rm lag})$. In this paper we specifically consider $\rho_\text{auto}=\langle S_1(t) S_1(t-t_{\rm lag})\rangle / \sigma^{2}_{1}$ and $\rho_\text{cross}=\langle S_1(t) S_2(t-t_{\rm lag})\rangle / \sigma_1 \sigma_2$, where $\sigma_1$ and $\sigma_2$ are the standard deviations of $S_1(t)$ and $S_2(t)$ in their steady states. The auto- and cross-correlation functions derived from a typical time trace of digitized SMTJ voltage states are shown in Figs.~\ref{fig:motivation_model}(b,c), showing damped oscillatory responses. Figures~\ref{fig:motivation_model}(d,e) show calculations of the correlation functions using a first-order Markov model as described in Sec.~\ref{sec:prob_corr_no_delay}. While this model reproduces the non-monotonic behavior found in the experimental data, it cannot reproduce the oscillation amplitude.  We attribute this failure of the model to the nonzero communication delay in the coupling between SMTJs. The circuits connecting the two SMTJs introduce delays of the order of microseconds; we denote the time delay from SMTJ $j$ to SMTJ $i$ as $t_{i \leftarrow j}$.

To demonstrate that the delay enhances the amplitude of the oscillations, we plot our theoretical result in Fig.~\ref{fig:motivation_model}(f,g), where the theory now accounts for the interaction delay (detailed results are provided in Sec.~\ref{sec:corr_delayed}) and leads to qualitative agreement with experiment. This result clearly indicates that models widely used in the stochastic computing community based on first-order Markov processes break down when interactions have significant delays, that is, when  the delay times  $t_{i \leftarrow j}$ are  longer than the intrinsic dwell times of each spin. Careful study is needed to understand the implications of such delays for Ising machine operation.

\section{Model} \label{sec:model}
In this section, we present our model of coupled Ising spins with time-delayed interactions. It consists of a probabilistic update rule for each individual Ising spin, accounts for time delays, and is based on a discretized approximation of time. 
Although this model is motivated by coupled SMTJs, it is relevant for any probabilistic system with delayed interactions.

An individual Ising spin $S_{i}$ has two possible states, spin up ($S_{i}=+1$) and spin down ($S_{i}=-1$). In the presence of thermal fluctuations, spins randomly flip between these two states. According to the N\'eel-Brown model~\cite{brown1963thermal}, the transition rate is $\lambda_{i}^{0}(T) = \lambda_{\text{flip}}^{0}\exp{(-\Delta E_{i}/kT)}$, with $\lambda_{\text{flip}}^{0}$, $\Delta E_{i}$, $k$, and $T$ the characteristic rate, the energy barrier between two states, the Boltzmann constant, and the temperature, respectively. When there is no field acting on the spin, the energy barriers for the $+1\rightarrow-1$ and $-1\rightarrow+1$ transitions are equal, and the transition rates between these two states is balanced. In contrast, if we apply an external or an effective field $B_{i}$, the energies of these two states become different, as the energy barriers are modified to $\Delta E_{i} + \mu S_{i}B_{i}$, where $\mu$ is the magnetic moment associated with each spin. Unbalanced transition rates lead to different probabilities of finding the spin in the $\pm1$ state when the spin reaches thermal equilibrium.

Interactions between individual spins, as well as onsite biases, can give rise to nonzero effective fields $B_{i}$. We denote the directed interaction from $S_{j}$ to $S_{i}$ as $J_{i\leftarrow j}$, and the bias field on $S_{i}$ as $h_{i}$ (in what follows, $J_{i\leftarrow j}$ and $h_{i}$ have physical dimensions of energy). Then we have $\mu B_{i} = \sum_{j\neq i} J_{i\leftarrow j} S_{j} + h_{i}$, and the transition rate becomes
\begin{equation} \label{eq:flipping_rate_no_delay}
    \lambda_{i} = \lambda_{i}^{0}(T)\exp{\left[-\frac{1}{k T}\left(\sum_{j\neq i} J_{i\leftarrow j} S_{j} + h_{i}\right)S_{i}\right]},
\end{equation}
where $\lambda_{i}^{0}(T)=\lambda_{\text{flip}}^{0}\exp{(-\Delta E_{i}/kT)}$.
If the interactions among all spins are fully symmetric, $J_{i\leftarrow j} = J_{j\leftarrow i} = J_{ij}$, then the Hamiltonian of the Ising model is well defined and it has the familiar form 
\begin{equation} \label{eq:Hamiltonian}
    H = - \sum_{i < j} J_{ij}S_{i}S_{j} - \sum_{i} h_{i}S_{i}.
\end{equation}
In the present paper, though, we are particularly interested in the implications of asymmetric coupling, \textit{namely} $J_{i\leftarrow j} \neq J_{j\leftarrow i}$, in which case there the non-Hermitian Hamiltonian does not in general yield a well-defined energy landscape.

We also allow interactions to be time-delayed such that it takes a time $t_{i\leftarrow j}$ for  information about the state of spin $S_{j}$ to arrive at, and influence, spin $S_{i}$. In other words, the transition rate of $S_{i}$ at time $t$ depends on the state of $S_{j}$ at time $(t - t_{i\leftarrow j})$. We modify Eq.~\eqref{eq:flipping_rate_no_delay} to represent this dependency, so that
\begin{multline} \label{eq:flipping_rate_delay}
    \lambda_{i}(t)= \lambda_{i}^{0}(T) \times  \\
    \exp{\left[-\frac{1}{k T}\left(\sum_{j\neq i} J_{i\leftarrow j} S_{j}(t - t_{i\leftarrow j}) + h_{i}\right)S_{i}(t)\right]}. 
\end{multline}
Equation \eqref{eq:flipping_rate_delay} is the starting point of our investigation. Note that, even if the coupling is fully symmetric, time-delayed interactions preclude a well-defined energy landscape just as a non-Hermitian Hamiltonian would.

We numerically simulate the dynamics corresponding to Eq.~\eqref{eq:flipping_rate_delay}. For systems with instantaneous interactions, Eq.~\eqref{eq:flipping_rate_no_delay} determines the flow of probability in the state space. Although such probability flows are well studied in continuous time via the Kolmogorov equations, here we discretize time and represent the probability distribution in a discrete form, which simplifies the treatment of our time-delayed interactions. The probability of spin $S_{i}$ to transition is approximately $\lambda_{i}\Delta t$, with $\Delta t$ the discretized time interval. We deploy a Monte Carlo algorithm to simulate transitions and system dynamics, choosing appropriate values of $\Delta t$, $\lambda^{0}_{\rm flip}$, $\Delta E_i$, and $J_{i\leftarrow j}$ such that  $\lambda_{i}\Delta t \ll 1$ and so that the probability of more than one spin transition within each time step is negligible. Furthermore, we constrain the magnitude of the coupling strength and bias fields such that the energy barriers of spins remain well-preserved---that is, such that they are subject to a Kramers-type analysis~\cite{kramers1940brownian}---ensuring the validity of the  exponential form of the transition rate.

\section{Steady Properties with Asymmetric Coupling} \label{sec:prob_corr_no_delay}
In this section, we study the steady-state probability distributions and correlation functions in the absence of time delay, particularly emphasizing the role of asymmetric coupling. To gain a clear understanding of system behavior without the complication of exponentially large state spaces, we consider only two coupled spins, with four total joint states, in Secs.~\ref{sec:prob_corr_no_delay} and~\ref{sec:prob_corr_delayed}. Although the two-spin system is greatly simplified, our main conclusions generalize to systems with more coupled spins, as discussed in Sec.~\ref{sec:generalizaiton}.

\subsection{Probability Distributions} \label{sec:prob_no_delay}
\begin{figure}[!htbp]
\centering
\includegraphics[width=1\linewidth]{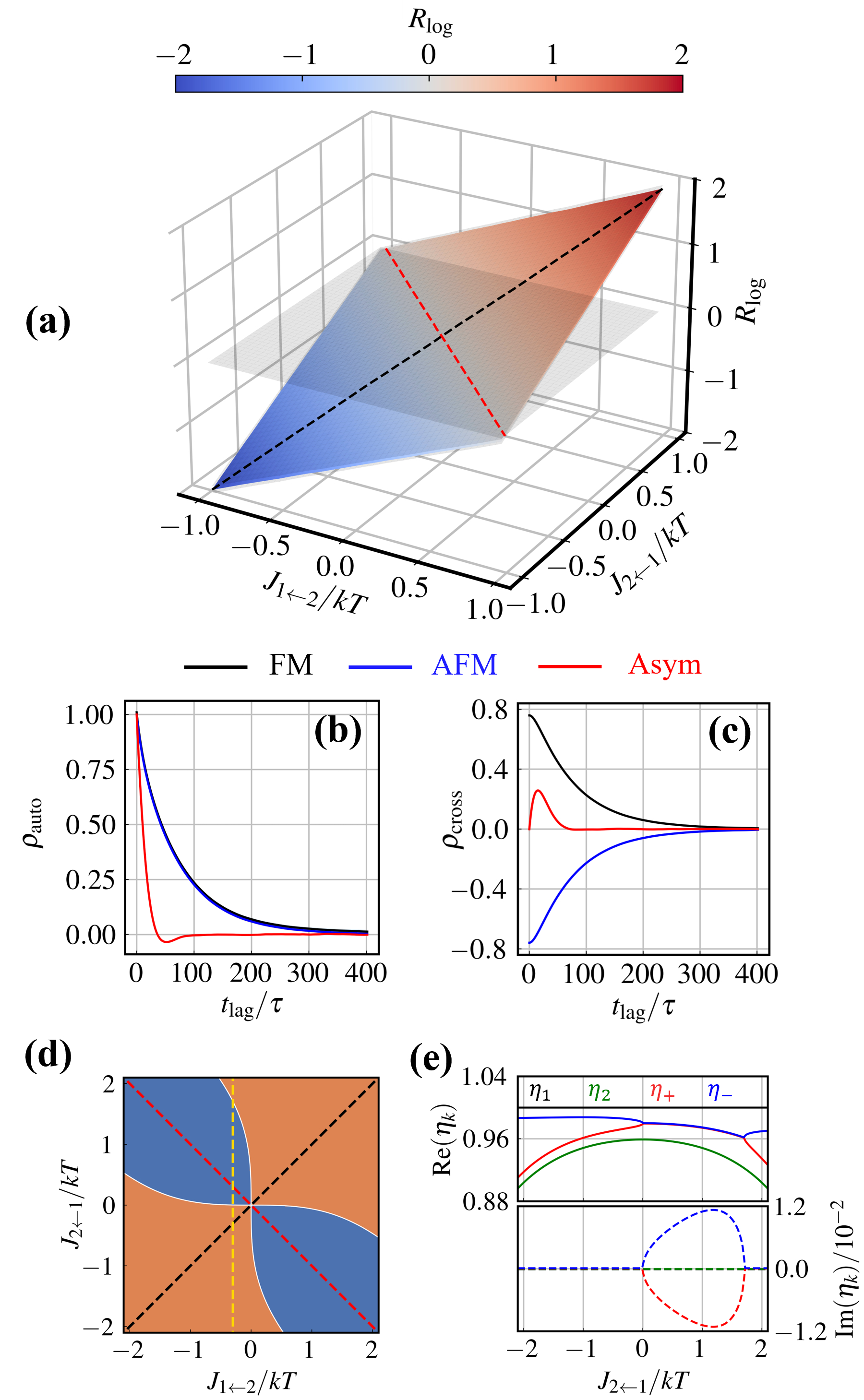}
\caption{ (a) Steady-state probability distribution as characterized by $R_{\rm log}$ as a function of interaction strengths. The auto- and cross-correlation functions are shown in (b,c) for ferromagnetically ($J_{1 \leftarrow 2} = J_{2 \leftarrow 1}>0$, FM), antiferromagnetically ($J_{1 \leftarrow 2} = J_{2 \leftarrow 1}<0$, AFM), and antisymmetrically ($J_{1 \leftarrow 2} = -J_{2 \leftarrow 1}>0$, Asym) coupled spins; $t_{\rm lag}$ is the time difference between two measurements of spin states and $\tau = 1/\lambda_{\rm flip}^{0}$ is the intrinsic characteristic time scale of each spin. (d) Phase diagram of oscillatory and non-oscillatory correlation functions. Blue and orange regions have the oscillatory and non-oscillatory correlation function depending on $J_{1 \leftarrow 2}$ and $J_{2 \leftarrow 1}$, respectively. Black and red dashed lines in (a,d) correspond to symmetric ($J_{1 \leftarrow 2} = J_{2 \leftarrow 1}$) and antisymmetric ($J_{1 \leftarrow 2} = -J_{2 \leftarrow 1}$) couplings. Real (solid lines) and imaginary parts (dashed lines) of eigenvalues $\eta_{k}$ along the yellow line in (d) are shown in (e). Different colors in (e) correspond to different $\eta_{k}$.  Overlapped lines in (b,e) are slightly shifted for visual clarity. (a)-(c) are simulated with $\tau = 4.98 \Delta t$, $\Delta E_{i} =4 k T$, $|J_{1 \leftarrow 2}| = |J_{2 \leftarrow 1}| = k T$ (for (b,c)), $h_1 = h_2 = 0$, $t_{1 \leftarrow 2} = t_{2 \leftarrow 1} = 0$, and are averaged over $10^{4}$ trials; after reaching steady state, each trial then collects statistical data over a period of $2500\times \Delta t$ in panel (a) and $5000\times\Delta t$ in panels (b,c). $95~\%$ confidence intervals of $R_{\rm log}$ and correlation functions are obtained from standard statistical analysis but are narrower than the width of the line. Panels (d,e) are calculated based on analytical expressions.}
\label{fig:no_delay_distrib_corr}
\end{figure}

When interactions are instantaneous, one can construct a first-order Markov model to describe the probabilistic dynamics of the system.
The elements of the transition matrix $\bm{T}$
are determined according to Eq.~\eqref{eq:flipping_rate_no_delay}. Diagonalizing $\bm{T}$ gives its eigenvalues $\eta_{k}$ and eigenvectors $\bm{r}_{k}$. The eigenvector associated with $\eta_{k}=1$ gives the steady-state distribution; all others have eigenvalues $|\eta_{k}| < 1$, and their contributions decay to zero in the long time limit. 
In the absence of bias (\textit{i.e.} taking $h_i = 0$) and assuming equal intrinsic rates $\lambda^{0}_{1}(T) = \lambda^{0}_{2}(T)$, this principal eigenvector yields steady-state probabilities
\begin{subequations} \label{eq:prob_nodelay}
    \begin{align}
        P(\downarrow\downarrow \text{or} \uparrow\uparrow) &= \frac{e^{(J_{1 \leftarrow 2}+J_{2\leftarrow 1})/k T}}{2[1+e^{(J_{1 \leftarrow 2}+J_{2\leftarrow 1})/k T}]} , \\
        P(\downarrow\uparrow \text{or} \uparrow\downarrow)& = \frac{1}{2[1+e^{(J_{1 \leftarrow 2}+J_{2\leftarrow 1})/k T}]}, 
    \end{align}
\end{subequations}
where $\downarrow\downarrow$ means $S_1 = -1$, $S_2 = -1$ with similar definitions for other states. Although the Hamiltonian of the system is not Hermitian when $J_{1\leftarrow 2}\neq J_{2 \leftarrow 1}$, the probability distribution still exponentially depends on $J_{1\leftarrow 2} + J_{2 \leftarrow 1}$, following a generalized Boltzmann distribution. Figure~\ref{fig:no_delay_distrib_corr}(a) plots the nonuniformity of the steady-state probability distribution. For two-spin systems, we characterize this nonuniformity through the log-ratio
\begin{align}
    R_{\rm log} = \ln\{[P(\uparrow \uparrow)+P(\downarrow \downarrow)]/[P(\downarrow \uparrow)+P(\uparrow \downarrow)]\}.
    \label{eq:rlog}
\end{align}
Here, using Eqs.~\eqref{eq:prob_nodelay},  $R_{\rm log} =  (J_{1 \leftarrow 2}+J_{2\leftarrow 1})/k T$  depends linearly on $J_{1 \leftarrow 2}$ and $J_{2\leftarrow 1}$, consistent with the flat surface in Fig.~\ref{fig:no_delay_distrib_corr}(a). The red line in Fig.~\ref{fig:no_delay_distrib_corr}(a), indicates the special case for fully antisymmetric coupling, $J_{1 \leftarrow 2} = -J_{2\leftarrow 1}$, for which all four states share the same probability and $R_{\rm log} = 0 $. In this case, the spins appear uncorrelated through their probabilities, even though they are coupled, as can be seen in other metrics like correlation functions.

\subsection{Correlation Functions} \label{sec:corr_no_delay}
While steady-state probability distributions only depend on a single eigenvector, correlation functions are determined by all eigenvectors with eigenvalues $|\eta_{k}|< 1$.
As a result, correlation functions reflect the spectral structure of $\bm{T}$ that cannot be gleaned from the steady-state probability distribution. We plot the simulated autocorrelation functions in Fig.~\ref{fig:no_delay_distrib_corr}(b). The definition of auto- and cross-correlation functions is given in Section~\ref{sec:motivation}.
For any two spins with symmetric coupling, whether ferromagnetically or antiferromagnetically coupled, the autocorrelation function decays without oscillation. 

In contrast, when the coupling strength is fully antisymmetric $J_{1\leftarrow2} = -J_{2 \leftarrow 1}$, the autocorrelation function exhibits a damped oscillation. This makes the behavior of antisymmetrically coupled spins distinct from that of uncoupled spins, even though there is no difference in the probability distribution. The oscillation is especially prominent in the simulated cross-correlation functions, plotted in Fig.~\ref{fig:no_delay_distrib_corr}(c). That figure shows that ferromagnetically coupled spins tend to be positively correlated, while antiferromagnetically coupled spins are negatively correlated, and in either case the correleation monotonically decays with $t_\text{lag}$. Asymmetrically coupled spins, however, exhibit in their cross-correlation a non-monotonic dependence on $t_{\rm lag}$. This reflects the oscillatory nature of correlations.

The oscillatory correlation functions, in the absence of bias, are determined by the eigenvectors of $\bm T$ as well as its eigenvalues, namely
\begin{subequations} \label{eq:T_eigval}
    \begin{align}
        \eta_{1} &= 1, \\
        \eta_{2} &= 1 - 2\lambda^{0} g\Delta t , \\
    \text{and}   \quad \eta_{\pm}& = 1 - \lambda^{0} (g \pm \sqrt{f})\Delta t  \nonumber,\\
        \text{where}\quad f &= (\cosh J_{1\leftarrow 2} - \cosh J_{2\leftarrow 1})^{2} \nonumber\\
        &\quad\quad\phantom{x}+ 4 \sinh J_{1\leftarrow 2} \sinh J_{2\leftarrow 1}, \\
        \text{and}\quad g &= \cosh J_{1\leftarrow 2} + \cosh J_{2\leftarrow 1} . 
    \end{align}
\end{subequations}
Here we assume $\lambda^{0} = \lambda^{0}_{i}(T)$ for $i=1,2$. The eigenvalues $\eta_{\pm}$ have non-zero imaginary part if $f<0$. In this case, since the correlation function has time dependence proportional to $(\eta_{k})^{n_{\rm lag}}$,
the contributions from $\eta_{+}$ and $\eta_{-}$ contribute to a time dependence of roughly  $|\eta_{\pm}|^{t_{\rm lag}/\Delta t}\cos (\theta_{\pm}t_{\rm lag}/\Delta t)$, where $\theta_{\pm}$ is the phase of $\eta_{\pm}$. Now we apply this qualitative analysis to the cases shown in Fig.~\ref{fig:no_delay_distrib_corr}(b). When the coupling strengths are symmetric ($J_{1\leftarrow2} = J_{2\leftarrow1} = J$),  we have $f = 4 \sinh^{2}J \ge 0$; if the coupling strengths are fully antisymmetric ($J_{1\leftarrow2} = -J_{2\leftarrow1} = J$), then $f = -4 \sinh^{2}J \le 0$. Therefore, the asymmetrically coupled spins acquire an oscillatory $\cos (\theta_{\pm}t_{\rm lag}/\Delta t)$ in the autocorrelation function while the symmetrically coupled spins do not, which conclusions explain the curves in Fig.~\ref{fig:no_delay_distrib_corr}(b). A similar qualitative analysis can be applied to the cross-correlation functions by replacing $\cos (\theta_{\pm}t_{\rm lag}/\Delta t)$ with $\sin (\theta_{\pm}t_{\rm lag}/\Delta t)$.

To understand what combinations of $J_{1\leftarrow2}$ and $J_{2\leftarrow1}$ lead to oscillatory or non-oscillatory behavior of correlation functions, we plot the phase diagram in Fig.~\ref{fig:no_delay_distrib_corr}(d). The phase boundary is determined by $f=0$ in Eqs.~\eqref{eq:T_eigval}. As expected, symmetric (antisymmetric) couplings always lead to non-oscillatory (oscillatory) correlation functions as shown by the black (red) lines in Fig.~\ref{fig:no_delay_distrib_corr}(d). For generically asymmetric couplings, holding one coupling strength ($J_{1\leftarrow2}$) constant while varying the other ($J_{2\leftarrow1}$) shifts the system passes into and then back out of the oscillatory regime (\textit{e.g.} along the yellow line). We visualize the real and imaginary parts of $\eta_{k}$ for this case in Fig.~\ref{fig:no_delay_distrib_corr}(e). We see that, in addition to the always-real $\eta_{1}$ and $\eta_{2}$, $\eta_{+}$ and $\eta_{-}$ are in general complex: when they are real in the non-oscillatory region, their values are distinguished; when they become non-real in the oscillatory region, they are complex conjugates with imaginary parts two orders smaller than their real part. These comparatively small imaginary parts explain why the correlation function for antisymmetric couplings in Fig.~\ref{fig:no_delay_distrib_corr}(b) appears strongly damped. This observation is robust against different values of $J_{1\leftarrow2}$ and $J_{2\leftarrow1}$. To see why, approximate $|\eta_{\pm}|^{t_{\rm lag}/\Delta t}\cos (\theta_{\pm}t_{\rm lag}/\Delta t)$ as $e^{-\lambda^{0} g t_{\rm lag}} \cos(\lambda^{0} \sqrt{|f|}t_{\rm lag})$ in the limit of $\Delta t \rightarrow 0$. The ratio of the oscillation frequency to the damping rate, $\sqrt{|f|} / g$, is bounded by unity regardless of $J_{1\leftarrow 2}$ and $J_{2\leftarrow 1}$.  For example, when $J_{1\leftarrow 2} = -J_{2\leftarrow 1} = J$,$\sqrt{|f|} / g =  \tanh |J| < 1$. 

This bound on $\sqrt{|f|}/g$ guarantees that oscillations will be strongly damped when the system is modeled by instantaneous interactions. It follows that the instantaneous-interaction assumption is incommensurate with the significant oscillations observed in the experiment, namely in the correlation functions in Fig.~\ref{fig:motivation_model}(b,c). To explain these experimentally observed oscillations, we now turn to the inclusion of time-delayed coupling in our model.

\section{Dynamics and Steady Properties with Finite Time Delay} \label{sec:prob_corr_delayed}
In this section, we add time delay to the coupling and investigate the dynamical evolution of both the probability distributions (Sec.~\ref{sec:prob_delayed}) and the steady-state correlation functions (Sec.~\ref{sec:corr_delayed}). In Sec.~\ref{sec:proof_delayed} we prove that, for two coupled spins, the probabilities of all four joint states are equal in the presence of time delays that are long compared to the intrinsic time scale of each spin. While the complete suppression of non-uniformity is exact for two coupled spins, for more than two spins this suppression is not total---we discuss the residual higher-order effects in Sec.~\ref{sec:generalizaiton}. Appendix~\ref{sec:Zn_spin_proof}, generalizes our conclusions concerning the suppression of non-uniformity to any stochastic network with long transmission delays that respects certain symmetry conditions, making it applicable in a wide range of systems including Potts models, Kuramoto models, Heisenberg models, nonlinear $\sigma$ models, and many engineered systems such as phase-based oscillator networks and complex-valued Ising machines. One noteworthy example is the suppressed collective frequency in the delayed stochastic Kuramoto model, where our result provides a general perspective on the previously known result~\cite{niebur1991collective}.

\subsection{Dynamical Evolution of Probability Distributions} \label{sec:prob_delayed}
\begin{figure}[!htbp]
\centering
\includegraphics[width=0.97\linewidth]{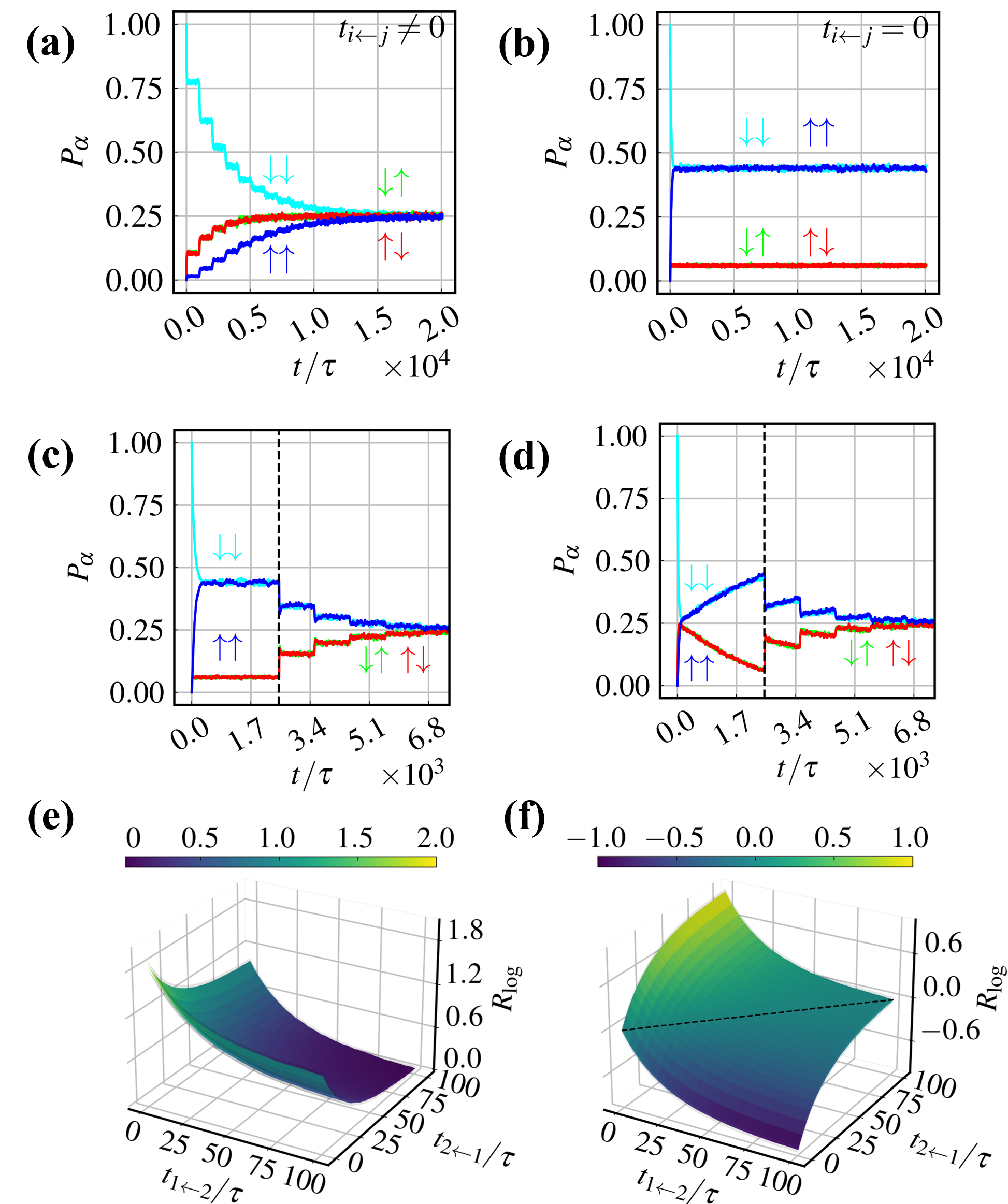}
\caption{Evolution of probability distribution of ferromagnetically coupled spins ($J_{1 \leftarrow 2} = J_{2 \leftarrow 1}>0$), where initial states are $\downarrow\downarrow$ for $t\le0$, with long time delay and without time delay shown in (a,b), respectively. In (c,d), two spins initially evolve without time delay and with constant $J_{1 \leftarrow 2}(t) = J_{2 \leftarrow 1}(t) = k T$ in (c) or linearly increasing $J_{1 \leftarrow 2}(t) = J_{2 \leftarrow 1}(t) = k T t/2510\tau$ in (d) until $t = 2510\tau$ (indicated by the dashed line). After that, two spins evolve with long time delay and constant $J_{1 \leftarrow 2}(t) = J_{2 \leftarrow 1}(t) = k T$. $t$ is the evolving time. Steady-state distribution of spin states, reflected by $R_{\rm log}$, as a function of time delay $t_{1 \leftarrow 2}$ and $t_{2 \leftarrow 1}$ with ferromagnetic and antisymmetric coupling ($J_{1 \leftarrow 2} = -J_{2 \leftarrow 1}>0$) shown in (e,f), respectively. The color bar and height of the surface correspond to $R_{\rm log}$. Black dashed line in (f) corresponds to $t_{1 \leftarrow 2} = t_{2 \leftarrow 1}$. $\tau = 1/\lambda_{\rm flip}^{0}$ is the intrinsic characteristic time scale of each spin in (a)-(f). Simulations are performed with $\tau = 4.98 \Delta t$, $\Delta E_{i} = 4 k T$, $|J_{1 \leftarrow 2}| = |J_{2 \leftarrow 1}| = k T$, $h_1 = h_2 = 0$, $t_{1 \leftarrow 2} = t_{2 \leftarrow 1} = 5000\Delta t$ for (a) and for $t>2510\tau$ in (c,d), and $t_{1 \leftarrow 2} = t_{2 \leftarrow 1} = 0$ for (b) and for $t<2510\tau$ in (c,d). Each data point gives the mean value averaged over $10^{4}$ ensembles for (a)-(f), $5000\Delta t$ period of time for (e,f) after reaching steady state. $95~\%$ confidence intervals of probabilities and $R_{\rm log}$ are obtained from standard statistical analysis but are narrower than the width of the line, for example, the  $95~\%$ confidence interval for $R_{\rm log}$ is $ \pm 0.057$.
}
\label{fig:delayed_no_bias_prob}
\end{figure}

The introduction of time delays fundamentally changes the probability distribution. We plot the time-dependent probability distribution of two ferromagnetically coupled ($J_{1\leftarrow2} = J_{2\leftarrow 1} >0$) spins with time delay much longer than the intrinsic time of spins, obtained from Monte Carlo simulations, in Fig.~\ref{fig:delayed_no_bias_prob}(a). There are two striking features: first, the evolution of probabilities is no longer smooth. Rather, plateaus emerge, and within each plateau, the probabilities are roughly constant. Second, although two spins are still coupled, the steady-state distribution in the long time limit is uniform with $P_{\alpha} = 1/4$ where $\alpha$ labels the joint state, as it would be if the two spins were uncoupled. In comparison, when the same coupling is not delayed, the steady-state distribution shown in Fig.~\ref{fig:delayed_no_bias_prob}(b) depends on the coupling strengths. 

An intuitive understanding of the plateaus in Fig.~\ref{fig:delayed_no_bias_prob}(a) is that until the system reaches steady state, the dynamics depends strongly on the initialization procedure. When $t_{1\leftarrow2} = t_{2\leftarrow1}$, the joint state at $t+1$ only depends on the state at $t$ and $t-t_{1\leftarrow 2}$. If we initialize both spins as $S_{i} = -1$ for $t<0$, then as the system starts to evolve until $t >t_{1\leftarrow 2}$, each spin can only see its neighbor as $S_{i} = -1$, and the two spins independently achieve their own equilibrium. This is the cause of the first plateau. When $t_{1\leftarrow2} < t < 2t_{1\leftarrow2}$, each spin sees its neighbor's state in the first plateau, whose statistical average is constant. As a result, each spin again quickly reaches its local equilibrium and the second plateau arises. The same process repeats until the system reaches a steady state. Similar plateau-like structures  were observed in long-delayed feedback optical systems, though regarding the laser intensity rather than probability distribution~\cite{giacomelli2012coarsening, giacomelli2013nucleation}.  

To test the idea that the appearance of plateaus strongly depends on the initialization procedure, we change the initialization, allowing the coupled spins to evolve with fixed coupling strength (Fig.~\ref{fig:delayed_no_bias_prob}(c)) or gradually increasing coupling strength (Fig.~\ref{fig:delayed_no_bias_prob}(d)) in the absence of delay. After a sufficient initialization period, we set the delay time $t_{1\leftarrow 2}=t_{2\leftarrow 1}>0$ and let the system evolve. If the distribution is constant before $t=2510~\tau$ as in Fig.~\ref{fig:delayed_no_bias_prob}(c), we find plateaus after $t=2510~\tau$. In contrast, when the coupling varies during the initialization period, as shown in Fig.~\ref{fig:delayed_no_bias_prob}(d), immediately after $t=2510~\tau$, the probability distribution follows the time dependence of initialization, so the curve still shows steps but is no longer flat between steps. In each of the subsequent time intervals, the distribution function follows the time dependence in the preceding time interval. In addition to the changed shape  of the plateaus, the time the system takes to reach steady state is shorter when the initial condition is not held constant. All these observations suggest that we can control the dynamics of the probability distribution by tuning the initial conditions before the delayed interaction is turned on.

To gain further insight into the conditions in which the spins seem uncoupled, we plot the dependence of steady-state distribution on the time delay, shown in Fig.~\ref{fig:delayed_no_bias_prob}(e) and (f). If $t_{1\leftarrow 2} = t_{2\leftarrow 1}$ and the two spins are coupled ferromagnetically, increasing the delay time gradually reduces the probability difference among the four states. As the probability distribution becomes uniform, $R_\text{log}\rightarrow 0$. For antiferromagnetically coupled spins (not plotted), the same trend holds. For two spins coupled fully asymmetrically, indicated by the black line in Fig.~\ref{fig:delayed_no_bias_prob}(f), the probabilities do not change and it is always $P_{\alpha} = 1/4$. However, if we set the delay times differently, $t_{1\leftarrow 2} \neq t_{2\leftarrow 1}$, the steady-state distribution is still non-uniform as seen in the corners where $t_{1\leftarrow 2} = 0$ and $t_{2\leftarrow 1} = 100~\tau$, or $t_{1\leftarrow 2} = 100~\tau$ and $t_{2\leftarrow 1} = 0$ in Fig.~\ref{fig:delayed_no_bias_prob}(e). More interestingly, in Fig.~\ref{fig:delayed_no_bias_prob}(f), by fixing one delay time to zero and increasing another delay time, we can change the two spins from seemingly uncoupled to statistically parallel or antiparallel to each other. 

Although the appearance of plateau-like dynamics requires the delay times to be at least one order of magnitude larger than the mean dwell time $1/\lambda_{i}^{0}(T)$ of each spin, the uniform distribution with a finite coupling does not require such large time delays. If the minimum delay time is around two times the mean dwell time of each spin, the probability distribution is already uniform. Even when the delay time is comparable to the mean dwell time, the probability distribution noticeably deviates from the one without delay. These results suggest that taking the time delay into consideration may be important for large-scale Ising machines designed for devices with fast intrinsic times and finite communication times.

\begin{figure}[!htbp]
\centering
\includegraphics[width=0.95\linewidth]{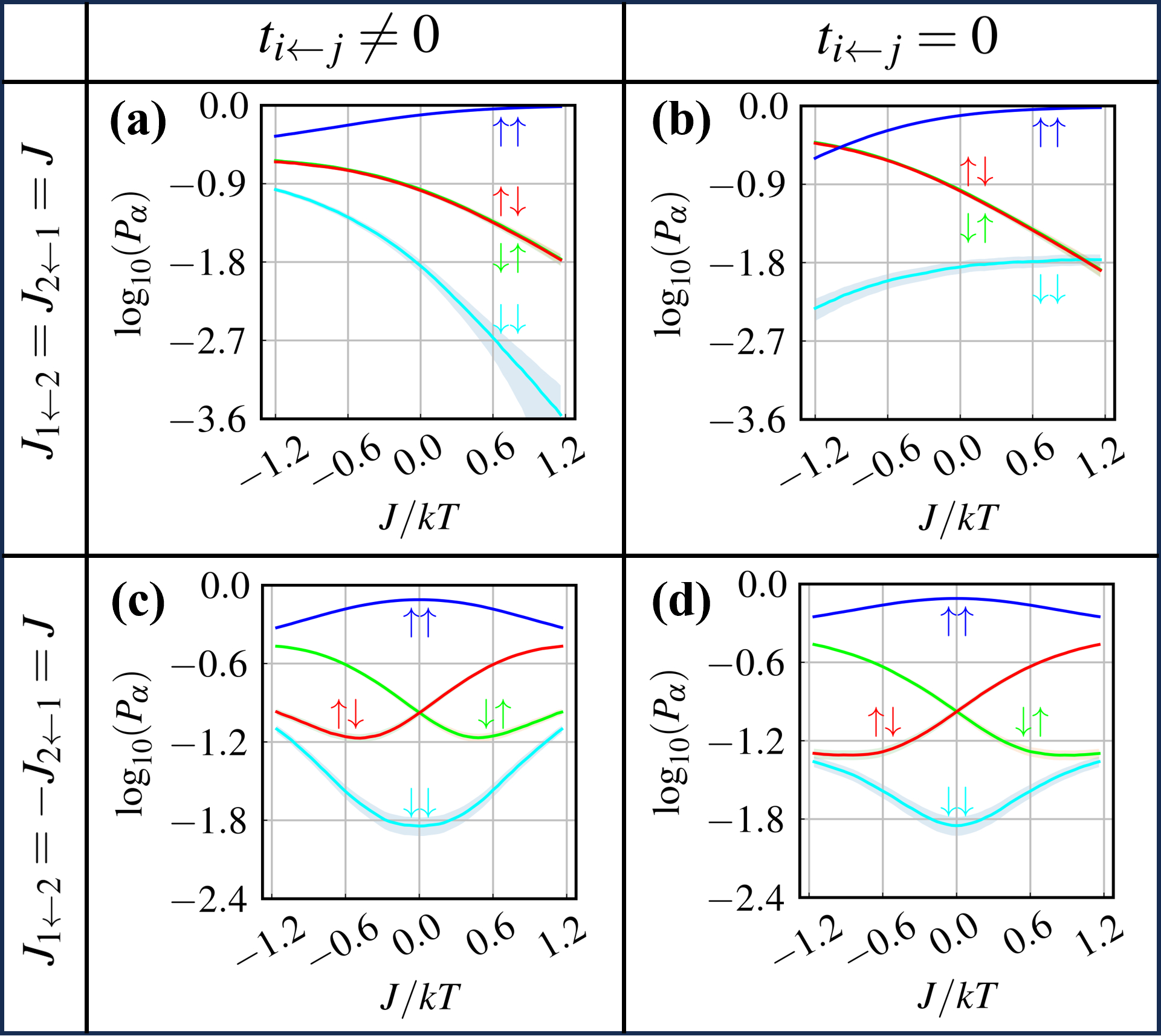}
\caption{Probability distributions in the presence of bias and time delay. Spins are coupled with symmetric coupling in (a,b), or with antisymmetric coupling in (c,d). There exists time delay $t_{1 \leftarrow 2} = t_{2 \leftarrow 1} = 500\Delta t$ in (a,c), or there is no time delay $t_{1 \leftarrow 2} = t_{2 \leftarrow 1} = 0$ in (b,d). Bias is $h_1 = h_2 = k T$ in (a)-(d). Overlapped lines are slightly shifted for visual clarity. Simulations are performed with $\tau = 4.98 \Delta t$, $\Delta E_{i} = 4k T$. Each data point gives the mean value averaged over $10^{4}$ ensembles and $5000\Delta t$ period of time after reaching steady state. $95~\%$ confidence intervals of $\log_{10}(P_{\alpha})$ obtained from standard statistical analysis are plotted as shaded regions.
}
\label{fig:delayed_bias_prob}
\end{figure}

When we introduce bias into the system, the steady-state probability distribution becomes complicated. Figure~\ref{fig:delayed_bias_prob} compares the steady-state distribution with symmetric or asymmetric coupling and with zero or non-zero time delay in the presence of a symmetric bias $h_1 = h_2 = kT$. We can see in Fig.~\ref{fig:delayed_bias_prob} (a) and (c) that the probability distribution is no longer uniform and depends on how the two spins are coupled. Furthermore, comparing Fig.~\ref{fig:delayed_bias_prob}(a) with (b) or (c) with (d), shows that in the presence of bias, time delay plays a role in determining the probability distribution, although this role is more prominent if the spins are strongly coupled and are not coupled fully antisymmetrically.

\subsection{Correlation Functions} \label{sec:corr_delayed}
\begin{figure}[!htbp]
\centering
\includegraphics[width=0.87\linewidth]{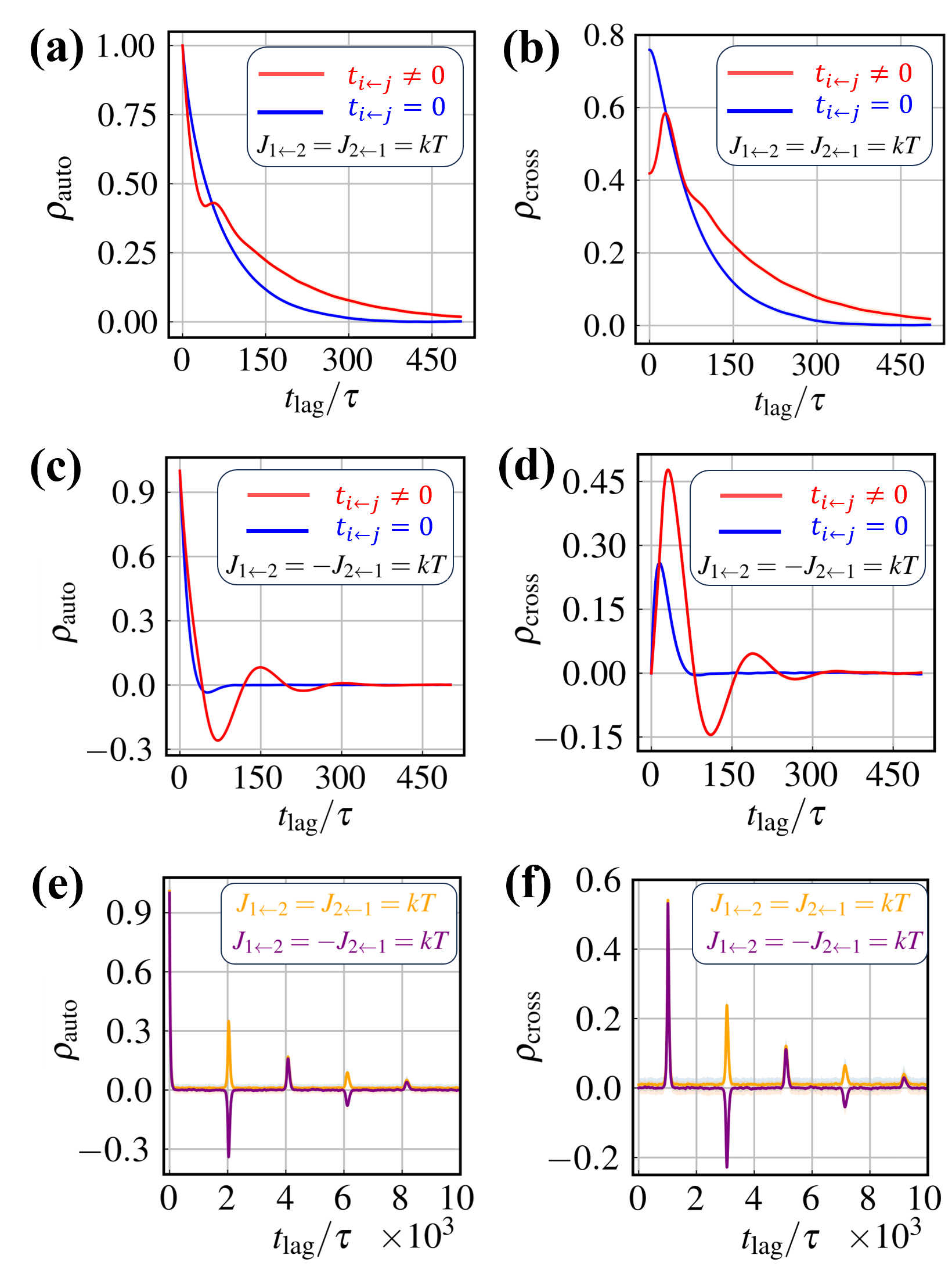}
\caption{Correlation functions of two coupled spins with short time delay, long time delay and without time delay. Spins are coupled ferromagnetically in (a,b,e,f), and antisymmetrically in (c)-(f). Auto- and cross-correlation functions are shown in (a,c,e) and (b,d,f), respectively. The time delay in (a)-(d) is short as $t_{1 \leftarrow 2} = t_{2 \leftarrow 1} = 100\Delta t$ for $t_{i \leftarrow j } \neq 0$ lines, and is long in (e,f) as $t_{1 \leftarrow 2} = t_{2 \leftarrow 1} = 5000\Delta t$. Overlap lines in (e,f) are slightly shifted for visual clarity. $t_{\rm lag}$ is the time difference between two measurements of spin states and $\tau = 1/\lambda_{\rm flip}^{0}$ is the intrinsic characteristic time scale of each spin. Simulations are performed with $\tau = 4.98 \Delta t$, $\Delta E_{i} =4 k T$. Each data point gives the mean value averaged over $10^{4}$ ensembles and $5000\Delta t$ period of time for (a)-(d) or 400 time steps for (e,f) after reaching steady state. $95~\%$ confidence intervals of correlation functions are obtained from standard statistical analysis for (a)-(f) but are narrower than the width of the line in (a)-(d).
}
\label{fig:delayed_nobias_corr}
\end{figure}

As we have seen in Sec.~\ref{sec:corr_no_delay}, although the probability distribution is uniform and the two spins are seemingly uncoupled with respect to the probability distribution, the correlation functions can still exhibit unexpected features. We plot them in Fig.~\ref{fig:delayed_nobias_corr}(a)-(d) with the delay time $t_{1\leftarrow 2} = t_{2\leftarrow 1} = 20.08~\tau$. If the two spins are coupled ferromagnetically, compared to the system without delay, the autocorrelation function has a peak on top of the exponential decay, as shown in Fig.~\ref{fig:delayed_nobias_corr}(a). This peak becomes more prominent when we look at the cross-correlation function shown in Fig.~\ref{fig:delayed_nobias_corr}(b). Both correlation functions decay more slowly than those without delay. If the two spins are coupled antisymmetrically, the changes in the correlation functions are more significant as shown in Fig.~\ref{fig:delayed_nobias_corr}(c) and (d). We can see that the autocorrelation function exhibits stronger oscillations, with three peaks compared to only one peak without time delay. The absolute value of the correlation is also significantly enhanced. The enhancement of the oscillations and correlation becomes clearer in the cross-correlation function, suggesting that compared to autocorrelation functions, cross-correlation functions can offer more insights into system behavior. Such enhancement of correlation and the existence of more than one peak agree quite well with the experimental observation presented in Fig.~\ref{fig:motivation_model}(b) and (c) for the antisymmetrically coupled SMTJs.

To understand the origin of the enhanced oscillations, we increase the delay time significantly, $t_{1\leftarrow 2} = t_{2 \leftarrow 1} = 1004~\tau$, and plot the correlation functions in Fig.~\ref{fig:delayed_nobias_corr}(e) and (f). In this case, the decay time of correlation functions is much shorter than the delay time, which effectively separates the intrinsic dynamics of each spin and the delay time of the coupling. The overall decaying feature is significantly slowed, and the delay-induced peaks are more prominent. The location of these peaks provides important information: for either ferromagnetic or antisymmetric coupling, the peaks of autocorrelation functions are at $2 n t_{i\leftarrow j}$, and the peaks of cross-correlation functions are at $(2 n + 1) t_{i\leftarrow j}$, for integers $n\geq0$. For the autocorrelation function, the information of spin $S_{i}(t)$ travels to its neighbor with $t_{j \leftarrow i}$ then comes back with $t_{i \leftarrow j}$ and correlates with $S_{i}(t+t_{i \leftarrow j} + t_{j\leftarrow i})$, leading to the first peak. This round trip takes a time $t_{i \leftarrow j} + t_{j\leftarrow i} = 2t_{i\leftarrow j}$. All additional round trips take even multiples of of $t_{i\leftarrow j}$. In comparison, for the cross-correlation functions, the information of spin $S_{i}(t)$ travels to its neighbor with $t_{j \leftarrow i}$ and correlates with $S_{j}(t + t_{j \leftarrow i})$, leading to the first peak. Each additional peak requires an additional round trip from $S_{j}$, so the the peaks of cross-correlation functions appear for odd multiples of $t_{j \leftarrow i}$. 
To make further connection with our experiment, where the current hardware implementation makes changing the delay time difficult, we show the dependence of correlation functions on the strength of antisymmetric coupling in Appendix~\ref{sec:animation}, which connects our theory to the experiments varying the coupling strength rather than the delay time. As coupling strength increases for a fixed delay, oscillations and peaks in the correlation functions appear.

\subsection{Theoretical Analysis of Suppressed Non-uniformity of Probability Distribution With Long Time Delay} \label{sec:proof_delayed}  

In Sec.~\ref{sec:prob_delayed}, we show that in the absence of bias and in the presence of long delay times, the non-uniformity of probability distribution is strongly suppressed. Specifically, the probability distribution of four joint states becomes uniform for two spins. Here we theoretically analyze this result in the case of two coupled spins. We start with the calculation of the cross-correlation function of two coupled spins without bias and with identical mean dwell time $1/\lambda^{0}_{i}(T) = 1/\lambda^0$. According to Eq.~\eqref{eq:flipping_rate_delay}, the transition probability of spin $S_{i}$ is
\begin{align} \label{eq:double_condition_prob}
    &\text{Pr}[S_{i}(t+1) = -s| S_{i}(t) = s, S_{j}(t - t_{i \leftarrow j})] = \lambda_{i} \Delta t  \nonumber \\
      & = \lambda^{0}\Delta t \left[ \cosh\left( \frac{J_{i \leftarrow j}}{kT} \right) - sS_{j}(t - t_{i\leftarrow j}) \sinh\left( \frac{J_{i \leftarrow j}}{kT} \right)\right]
\end{align}
where $s=\pm 1$ and binarized values of $sS_{j}(t - t_{i\leftarrow j}) =\pm 1$ are utilized. Note that Eq.~\eqref{eq:double_condition_prob} is exact without requiring weak coupling expansion. This transition probability is invariant under both spins flipped, corresponding to $\mathbb{Z}_2$ symmetry. As a result, in the steady state, each spin has equal probability of two spin states, $P(S_{i}=\pm 1)=1/2$, and standard deviation of $S_{i}$ is $\sigma_{i} = 1$. We therefore have $\rho_{\rm cross}^{12}(t,t_{\rm lag}) = \left<S_{1}(t) S_{2}(t-t_{\rm lag}) \right>$ and $\rho_{\rm auto}^{ii}(t,t_{\rm lag}) = \left<S_{i}(t) S_{i}(t-t_{\rm lag}) \right>$. For equal-time cross-correlation function $\rho_{\rm cross}^{12}(t,0)$, we differentiate it with respect to $t$, and in the steady state limit, we have
\begin{align} \label{eq:2spin_cross_origin}
    &\frac{\partial \rho_{\rm cross}^{12}(t,0)}{\partial t} = -2 \sum_{i=1}^{2} \left< \lambda_{i}  S_{1}(t) S_{2}(t) \right> = 0 \nonumber \\
    &\Rightarrow \rho_{\rm cross}^{12}(t,0)   \nonumber \\
    & =\frac{\sinh \mathcal{J}_{1\leftarrow 2} \rho_{\rm auto}^{22}(t, t_{1 \leftarrow 2}) + \sinh \mathcal{J}_{2\leftarrow 1} \rho_{\rm auto}^{11}(t, t_{2 \leftarrow 1})}{\cosh \mathcal{J}_{1 \leftarrow 2} + \cosh \mathcal{J}_{2 \leftarrow 1}},
\end{align}
where $\mathcal{J}_{i \leftarrow j} \equiv J_{i \leftarrow j}/kT$. In order to find auto- and cross-correlation functions, in the steady state limit, we differentiate them with respect to $t_{\rm lag}$:
\begin{subequations} \label{eq:2spin_cross_auto_lag}
    \begin{align}
        &\frac{1}{2\lambda^0}\frac{\partial \rho_{\rm auto}^{ii}}{\partial t_{\rm lag}} = -\frac{1}{\lambda^0} \left< \lambda_{i}(t+t_{\rm lag}) S_{i}(t) S_{i}(t+t_{\rm lag}) \right> \nonumber \\
        &= - \cosh \mathcal{J}_{i \leftarrow j} \rho_{\rm auto}^{ii} + \sinh \mathcal{J}_{i \leftarrow j} \rho_{\rm cross}^{ij}(t, t_{i \leftarrow j} - t_{\rm lag}), \\
        &\frac{1}{2\lambda^0}  \frac{\partial \rho_{\rm cross}^{ij}}{\partial t_{\rm lag}}  = -\frac{1}{\lambda^0} \left< \lambda_{i}(t+t_{\rm lag}) S_{j}(t) S_{i}(t+t_{\rm lag}) \right> \nonumber \\
        & = -\cosh \mathcal{J}_{i \leftarrow j} \rho_{\rm cross}^{ij} + \sinh \mathcal{J}_{i \leftarrow j} \rho_{\rm auto}^{jj}(t, t_{i \leftarrow j} - t_{\rm lag}), 
     \end{align}
\end{subequations}
where we restrict $t_{\rm lag}$ within the interval of $0\leq t_{\rm lag} \leq \text{min}(t_{1\leftarrow 2}, t_{2\leftarrow 1})$. Eq.~\eqref{eq:2spin_cross_auto_lag} can be solved with the initial conditions specified by Eq.~\eqref{eq:2spin_cross_origin} and $\rho_{\rm auto}^{ii}(t,0) = 1$. To simplify the analytical expression, we assume uniform delay times $t_{1\leftarrow 2} = t_{2\leftarrow 1} = t_{d}$, and the equal-time cross-correlation function is
\begin{align} \label{eq:2spin_cross_orgin_uni_dealy}
    \rho_{\rm cross}^{12}(t,0) = \frac{\sinh \bar{\mathcal{J}} }{ \cosh(2 \lambda^{0} t_{d}) \cosh \bar{\mathcal{J}} + \sinh(2\lambda^{0} t_{d}) \cosh \Delta \mathcal{J} },   
\end{align}
where $\bar{\mathcal{J}}  \equiv (\mathcal{J}_{1\leftarrow 2} + \mathcal{J}_{2\leftarrow 1})/2$, $\Delta \mathcal{J}  \equiv (\mathcal{J}_{1\leftarrow 2} - \mathcal{J}_{2\leftarrow 1})/2$. 

Steady state probability distribution of four joint states is derived from the cross-correlation function according to the relation~\cite{donnell2014analysis}
\begin{equation} \label{eq:2spin_prob_cross_corr}
    P(s_{1}s_{2})=\frac{1}{4}\left[1 + \rho_{\rm cross}^{12}(t,0) s_{1}s_{2} \right],
\end{equation}
where $S_{i}(t) = s_{i}$. We can see that, regardless of symmetric or asymmetric $J_{i \leftarrow j}$, when $\lambda^{0} t_{d} \gg 1$, $\rho_{\rm cross}^{12}(t,0)\to 0$ so $P(s_{1}s_{2})\to1/4$. Even if delay times are non-uniform, as long as $ \lambda^{0} \text{min}(t_{1\leftarrow 2},t_{2\leftarrow 1}) \gg 1$ is satisfied, the same conclusion still holds, although the specific form of $\rho_{\rm cross}^{12}(t,0)$ is different from Eq.~\eqref{eq:2spin_cross_orgin_uni_dealy}. An intuitive argument is that each spin's information travels back to itself after a round trip and strongly correlates to itself, indicating the appearance of the first peak at $t_{\rm lag} \sim t_{1 \leftarrow 2} + t_{2 \leftarrow 1} > \text{max}(t_{1\leftarrow 2},t_{2\leftarrow 1})$. With the stated condition, $\rho_{\rm auto}^{ii}(t,t_{j \leftarrow i})$ takes its value after sufficient decay and before the appearance of first peak, therefore $\rho_{\rm auto}^{ii}(t,t_{j \leftarrow i}) \approx 0$. According to Eq.~\eqref{eq:2spin_cross_origin}, $ \rho_{\rm cross}^{12}(t,0)$ approaches 0, so $P(s_{1}s_{2})$ approaches $1/4$.

So far derivations do not require perturbative expansion of the transition rate $\lambda_{i}$. However, when we consider more coupled spins, we have to expand $\lambda_{i}$ in the weak coupling regime, and use perturbation theory to evaluate auto- and cross-correlation functions. Although in the long delay time limit, the probability distribution of joint spin states is not strictly uniform, its dependence on the coupling strength is strongly suppressed from first order to second order. The numerical demonstration is discussed in Sec.~\ref{sec:generalizaiton}, and detailed derivations are included in Appendix.~\ref{sec:corr_prob_time_deriv}. Furthermore, we generalize this conclusion and its proof from Ising systems, or $\mathbb{Z}_{2}$ spins, to broad classes of spin systems, including but not limited to $\mathbb{Z}_{n}$, $\text{SO(n)}$ and $\text{SU}(n)$ spins, which is included in Appendix.~\ref{sec:Zn_spin_proof}.

\section{Coupling Multiple Spins} \label{sec:generalizaiton}

The above discussions address steady-state distributions and correlation functions for two coupled spins with time delay. Here, we show that our central conclusions apply to systems with multiple spins. We demonstrate this generality with a system of five coupled spins.

\begin{figure}[!htbp]
\centering
\includegraphics[width=0.97\linewidth]{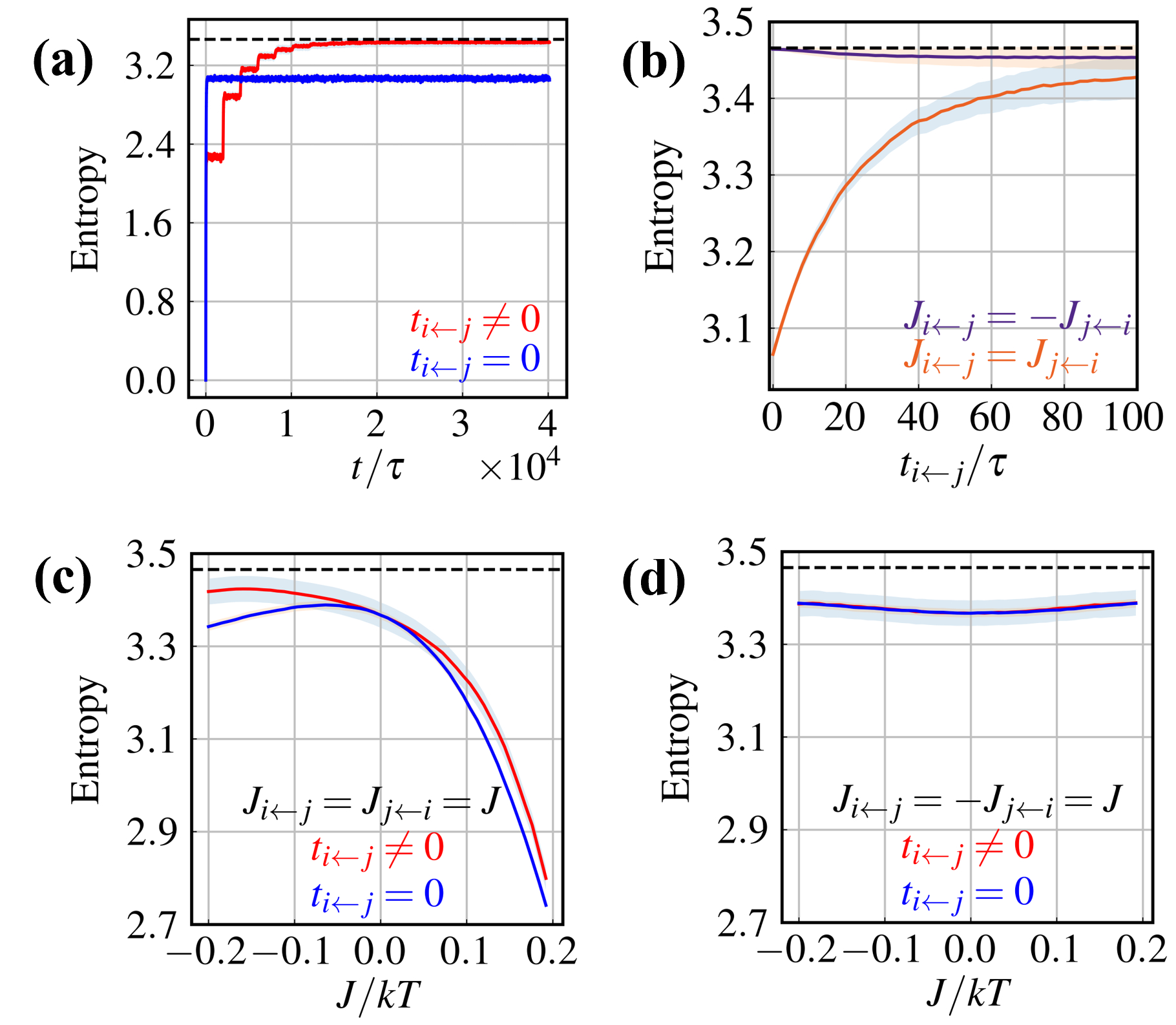}
\caption{Entropy of five coupled spins. (a) evolution of entropy with respect to time, $t$, of 5 ferromagnetically coupled spins ($J_{i \leftarrow j} = J_{j \leftarrow i} = 0.2 k T$), where initial states are $S_{i} = -1$, with long time delay and without time delay. (b) Dependence of entropy of 5 ferromagnetically ($J_{i \leftarrow j} = J_{j \leftarrow i} = 0.2 k T$) and antisymmetrically ($J_{i \leftarrow j} = -J_{j \leftarrow i} = 0.2 k T$) coupled spins on the time delay $t_{i \leftarrow j}$. Entropy dependence on the coupling strengths in the presence of bias is shown in (c) for symmetric coupling and (d) for antisymmetric coupling. In (a)-(d), black dashed lines correspond to the maximum possible entropy of $5\ln2$, and $t_{i \leftarrow j} = t_{j \leftarrow i}$. $\tau = 1/\lambda_{\rm flip}^{0}$ is the intrinsic characteristic time scale of each spin in (a,b). Simulations are performed with $\tau = 4.98 \Delta t$, $\Delta E_{i} = 4k T$, $h_i = 0$ for (a,b), $h_i = 0.2k T$ for (c,d), $t_{i \leftarrow j} = 10^{4}\Delta t$ for (a) and $t_{i \leftarrow j} = 500\Delta t$ for (c,d). Each data point gives the mean value averaged over $10^{4}$ ensembles, $5000\Delta t$ period of time for (b)-(d) after reaching steady state. $95~\%$ confidence intervals of entropy obtained from standard statistical analysis are plotted as shaded regions.
}
\label{fig:delayed_entropy_5SMTJ}
\end{figure}

For five coupled Ising spins, the system has 32 joint states, making it harder to visualize the probability distribution compared to two coupled spins. Since we are particularly interested in the uniformity of the distribution, we use the distribution's entropy, defined as $-\sum_{\alpha}P^{t}_{\alpha} \ln P^{t}_{\alpha}$, to quantify their uniformity. The maximum value of the entropy, (namely $N\ln 2$, for $N$ spins), is reached if and only if the distribution is uniform. We compare the evolution of entropy with a long time delay to that without delay in Fig.~\ref{fig:delayed_entropy_5SMTJ}(a). The entropy with delay has plateaus, and in the steady-state limit, its value is close to the maximum value, indicating the distribution has become nearly uniform. In comparison, when the five spins are coupled ferromagnetically without delay, the steady-state distribution is far from uniform. We compare the entropy dependence on the delay time for ferromagnetically and fully antisymmetrically coupled spins in Fig.~\ref{fig:delayed_entropy_5SMTJ}(b) in the absence of bias. Similar to the $t_{1\leftarrow2}=t_{2\leftarrow1}$ lines in Fig.~\ref{fig:delayed_no_bias_prob}(e) and (f), with increasing delay time, ferromagnetically coupled spins gradually become nearly uniformly distributed, while antisymmetrically coupled spins are always so within statistical error. When we add bias into the system, the $\mathbb{Z}_2$ symmetry is broken and the steady-state distribution depends on the coupling strengths, as shown in Fig.~\ref{fig:delayed_entropy_5SMTJ}(c) and (d). Such dependence is different for systems with and without delay, and for symmetrically coupled spins, the entropy is more sensitive to the coupling strengths than for antisymmetrically coupled spins. Since the calculation of entropy obscures the probability of each individual state, although each individual probability is not an even function and by flipping the sign of $J$ the values of probability shuffle among states, similar to Fig.~\ref{fig:delayed_bias_prob}(c) and (d), the entropy is an even function for asymmetrically coupled spins, as shown in Fig.~\ref{fig:delayed_entropy_5SMTJ}(d).

\begin{figure}[!htbp]
\centering
\includegraphics[width=0.95\linewidth]{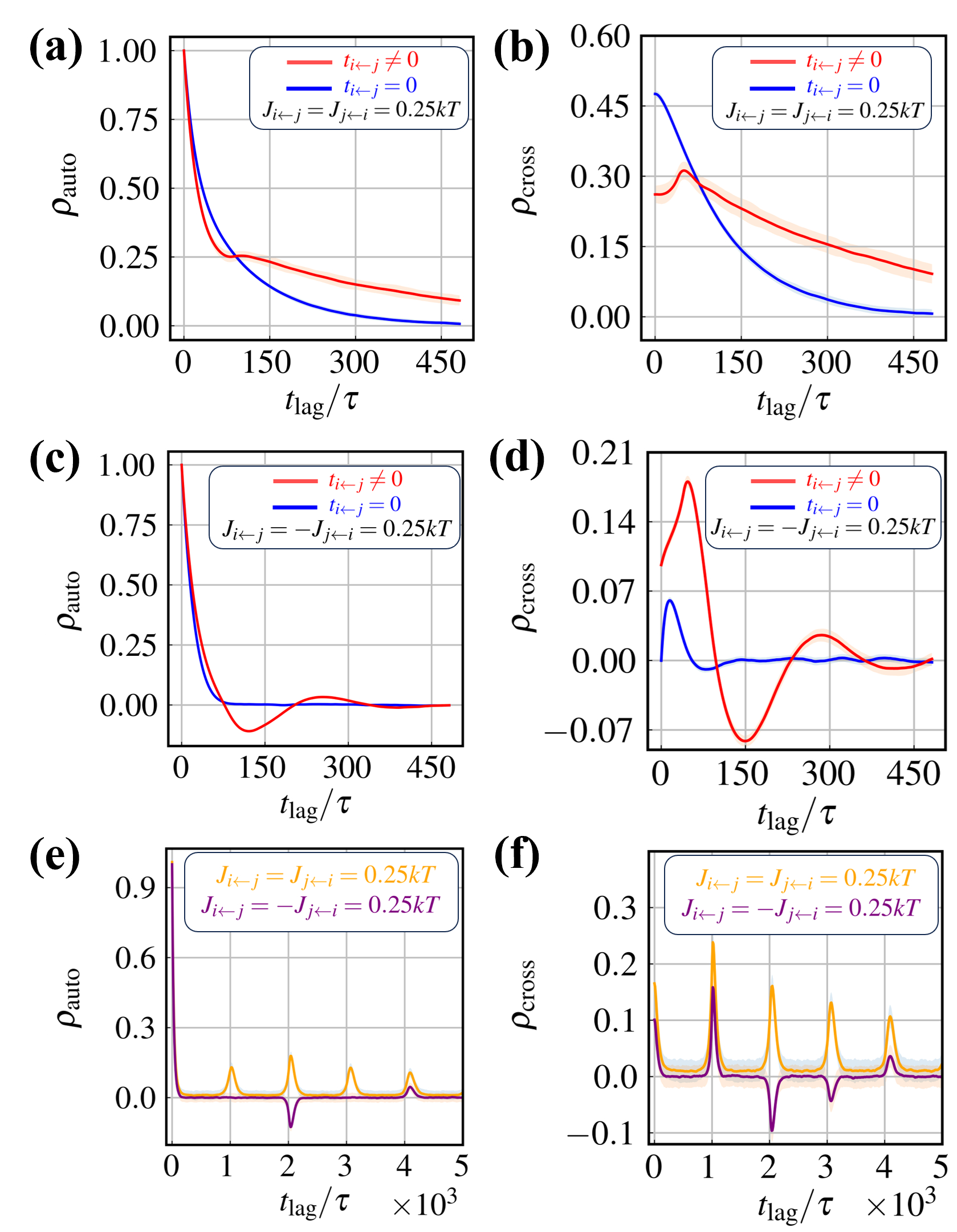}
\caption{Correlation functions of five coupled spins with short time delay, long time delay and without time delay. $\rho_{\rm cross}$ and $\rho_{\rm auto}$ are defined as $\rho_{\rm cross}^{12}$ and $\rho_{\rm auto}^{11}$, respectively. Spins are coupled ferromagnetically in (a,b,e,f), and antisymmetrically in (c)-(f). Auto- and cross-correlation functions are shown in (a,c,e) and (b,d,f), respectively. The time delay in (a)-(d) is short as $t_{i \leftarrow j} = t_{j \leftarrow i} = 200\Delta t$ for $t_{i \leftarrow j } \neq 0$ lines, and is long in (e,f) as $t_{i \leftarrow j} = t_{j \leftarrow i} = 5000\Delta t$. Overlap lines in (e,f) are slightly shifted for visual clarity. $t_{\rm lag}$ is the time difference between two measurements of spin states and $\tau = 1/\lambda_{\rm flip}^{0}$ is the intrinsic characteristic time scale of each spin. Simulations are performed with $\tau = 4.98 \Delta t$, $\Delta E_{i} = 4k T$. Each data point gives the mean value averaged over $10^{4}$ ensembles and $1250\Delta t$ period of time for (a)-(d) or 400 time steps for (e,f) after reaching steady state. $95~\%$ confidence intervals of correlation functions obtained from standard statistical analysis are plotted as shaded regions.
}
\label{fig:delayed_nobias_corr_5SMTJ}
\end{figure}

The correlation functions of five coupled spins, whose definitions follow those of two coupled spins, are plotted in Fig.~\ref{fig:delayed_nobias_corr_5SMTJ}. The correlation functions are oscillatory for fully antisymmetrically coupled spins even without time delay, weakly so for the autocorrelation and more pronounced for the cross-correlation. With finite time delay, the oscillations and correlations are both enhanced, most prominently for the cross-correlation function of antisymmetrically coupled spins, as was the case for the correlation functions of two coupled spins discussed above.

Figure~\ref{fig:delayed_nobias_corr_5SMTJ}(e,f) shows the correlation functions with long delay times. The multiple peak structure, observed in Fig.~\ref{fig:delayed_nobias_corr}(e,f) is preserved, but with subtle differences. For five coupled spins, the equal-time cross-correlation function, $\rho_{\rm cross}(0)$, has non-zero values for both ferromagnetic and antisymmetric couplings, which is not the case for two coupled spins. In addition, the peaks of autocorrelation functions, $\rho_{\rm auto}$, with ferromagnetic couplings occur at all integer delay times rather than only at even multiple delay times. The peaks of $\rho_{\rm auto}$ for antisymmetric couplings are at different locations compared to those for ferromagnetic couplings.

\begin{figure}[!htbp]
\centering
\includegraphics[width=1\linewidth]{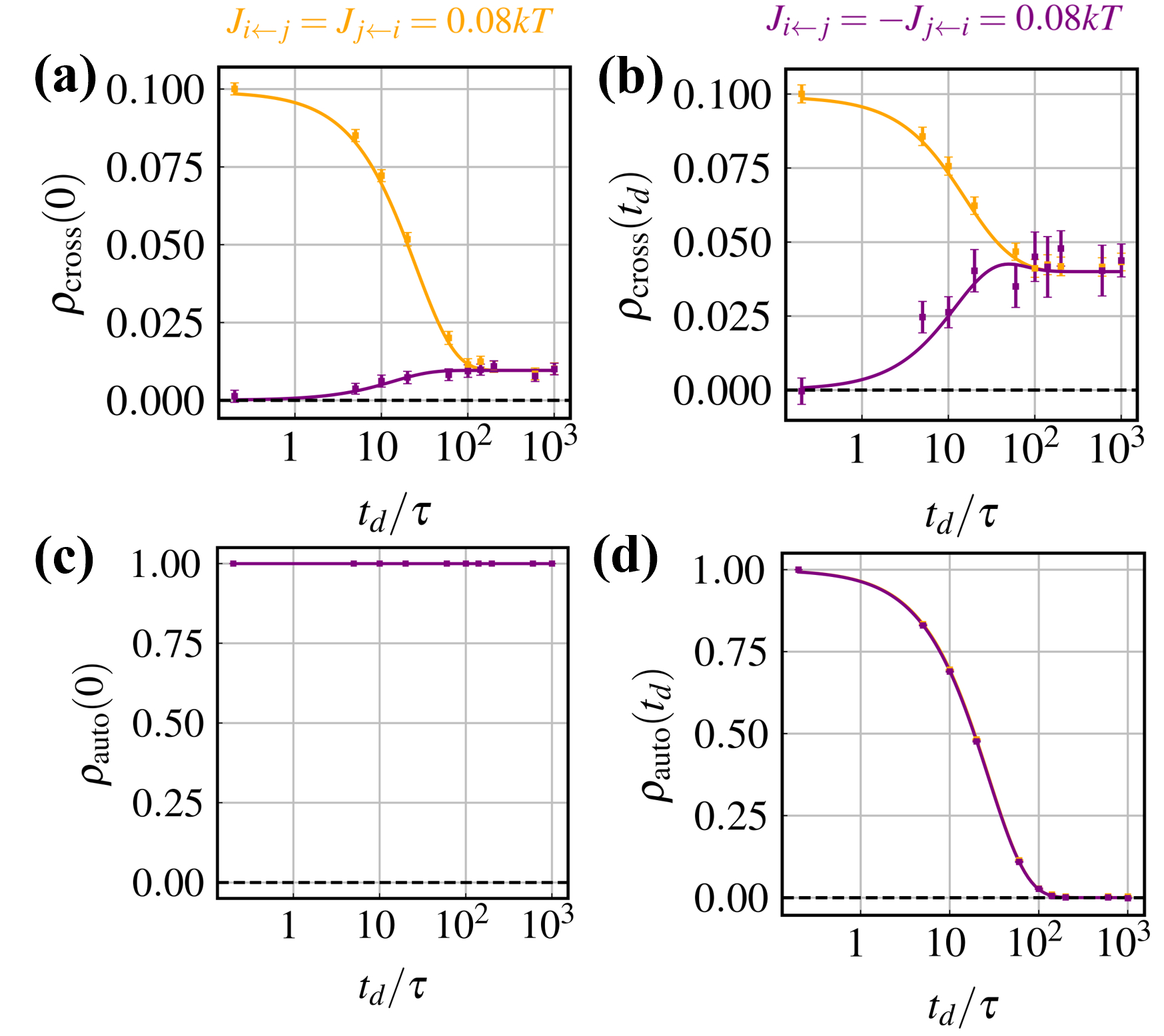}
\caption{Cross- and autocorrelation functions of five coupled spins at $t_{\rm lag}=0$ and $t_{\rm lag}=t_{d}$ with varying time delay $t_{d}$. $\rho_{\rm cross}(t_{\rm lag}) = \rho^{12}_{\rm cross}(t_{\rm lag})$ and $\rho_{\rm auto}(t_{\rm lag}) = \rho^{11}_{\rm auto}(t_{\rm lag})$, where $t_{\rm lag}$ is the time difference between two measurements of spin states. Delay times are uniform $t_{i \leftarrow j} = t_d$. Spins are coupled ferromagnetically ($J_{i \leftarrow j} = J_{j\leftarrow i} = 0.08 kT$) or antisymmetrically ($J_{i \leftarrow j} = -J_{j\leftarrow i} = 0.08 kT$). Solid lines are plotted based on Eqs.~\eqref{eq:main_1st_cross_origin}-\eqref{eq:main_2nd_cross_peak} for (a,b), $\rho_{\rm auto}^{(0)}(t_d)=e^{-2 \lambda^0 t_d}$ and Eq.~\eqref{eq:2nd_auto_peak} for (d). Simulations are performed with $\tau = 4.98 \Delta t$, $\Delta E_{i} =4 k T$. Each data point gives the mean value averaged over $10^{4}$ ensembles and 9975 $\Delta t$ periods of time after reaching steady state. $95~\%$ confidence intervals of correlation functions, plotted as error bars, are obtained from standard statistical analysis.
}
\label{fig:corr_peaks}
\end{figure}

\begin{figure}[!htbp]
\centering
\includegraphics[width=1\linewidth]{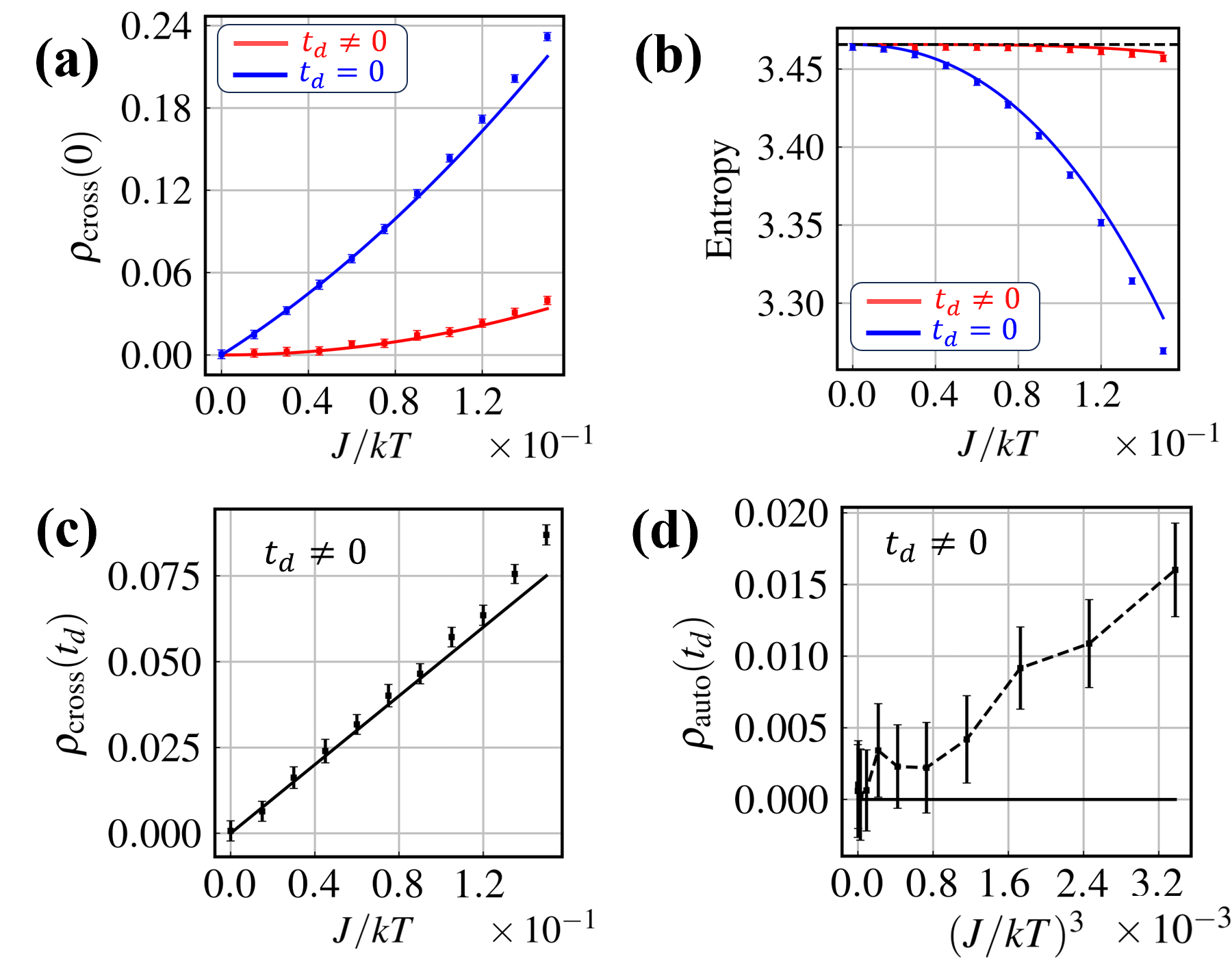}
\caption{Cross- and autocorrelation functions and entropy of five ferromagnetically coupled spins with varying coupling strengths and long time delay. $\rho_{\rm cross}(t_{\rm lag}) = \rho^{12}_{\rm cross}(t_{\rm lag})$ and $\rho_{\rm auto}(t_{\rm lag}) = \rho^{11}_{\rm auto}(t_{\rm lag})$, where $t_{\rm lag}$ is the time difference between two measurements of spin states. Delay times are uniform $t_{i \leftarrow j} = t_d$. Spins are coupled ferromagnetically ($J_{i \leftarrow j} = J_{j\leftarrow i} = J >0$), and delay times are $t_{d} = 5000\Delta t$ or $0$. In (b), the black dashed line corresponds to the maximum entropy of $5\ln2$. Solid lines are plotted based on Eqs.~\eqref{eq:main_1st_cross_origin}-\eqref{eq:main_2nd_cross_peak} for (a,c), \eqref{eq:main_1st_delta_P}-\eqref{eq:main_2nd_delta_P} for (b), $\rho_{\rm auto}^{(0)}(t_d)=e^{-2 \lambda^0 t_d}$ and Eq.~\eqref{eq:2nd_auto_peak} for (d), and the black dashed line in (d) simply connects data points. Simulations are performed with $\tau = 4.98 \Delta t$, $\Delta E_{i} =4 k T$. Each data point gives the mean value averaged over $10^{4}$ ensembles and $9975\Delta t$ period of time after reaching steady state. $95~\%$ confidence intervals of correlation functions, plotted as error bars, are obtained from standard statistical analysis.
}
\label{fig:corr_entropy_J}
\end{figure}

To understand the dependence of correlation functions on the delay time and coupling strength, we treat the case with uniform delay times ($t_{i \leftarrow j} = t_{d}$ for all $i,j$) and use perturbation theory in the weak coupling regime. Expanding the correlation function $\rho_{\rm cross}^{ij}$ in powers $n$ of the coupling strength gives the terms $\rho_{\rm cross}^{ij,(n)}(t_{\rm lag})$, starting with $\rho_{\rm cross}^{ij,(0)} = 0$  (see Appendix.~\ref{sec:corr_prob_time_deriv} for details),
\begin{subequations}
\label{eqs:cross-coors}
    \begin{align}
    &\rho_{\rm cross}^{ij,(1)}(0)  =\frac{e^{-2\lambda^{0} t_{d}}}{2}(\mathcal{J}_{i \leftarrow j} + \mathcal{J}_{j \leftarrow i}),\label{eq:main_1st_cross_origin} \\
    &\rho_{\rm cross}^{ij,(2)}(0) = \frac{1}{4}\sum_{k\neq i, j}  \left[ 2 \mathcal{J}_{i \leftarrow k}\mathcal{J}_{j \leftarrow k}  \right. \nonumber \\
    & \left. + (\mathcal{J}_{i \leftarrow k}\mathcal{J}_{k \leftarrow j} + \mathcal{J}_{j \leftarrow k} \mathcal{J}_{k \leftarrow i}) e^{-4 \lambda^{0} t_{d}} \right],  \label{eq:main_2nd_cross_origin} \\
    &\rho_{\rm cross}^{ij,(1)}(t_{d}) = \frac{\mathcal{J}_{i \leftarrow j} + \mathcal{J}_{j\leftarrow i}e^{-4\lambda^{0} t_{d}}}{2},  \label{eq:main_1st_cross_peak} \\
    &\rho_{\rm cross}^{ij,(2)}(t_{d}) = \frac{e^{-2 \lambda^0 t_{d}}}{4}\sum_{k\neq i, j}  \left[ (2 + 4\lambda^0 t_{d}) \mathcal{J}_{i\leftarrow k} \mathcal{J}_{j\leftarrow k} \right. \nonumber \\
    &\left. + \mathcal{J}_{j\leftarrow k} \mathcal{J}_{k\leftarrow i} e^{-4 \lambda^0 t_{d}} + \mathcal{J}_{i\leftarrow k} \mathcal{J}_{k\leftarrow j}  \right], \label{eq:main_2nd_cross_peak}
    \end{align}
\end{subequations}
where $\lambda^{0}_{i}(T) = \lambda^0$ and $\mathcal{J}_{i \leftarrow j} \equiv J_{i \leftarrow j}/kT$. Equations~\eqref{eq:main_1st_cross_origin} and \eqref{eq:main_2nd_cross_origin} show that delay suppresses the dependence of $\rho_{\rm cross}^{ij}(0)$ on the coupling strength from first order to second order in the long delay time limit. More specifically, for  $\lambda^0 t_{d} \gg 1$, $\rho_{\rm cross}^{ij,(1)}(0) \approx 0$ and $\rho_{\rm cross}^{ij,(2)}(0) \approx \sum_{k\neq i, j}  \mathcal{J}_{i \leftarrow k}\mathcal{J}_{j \leftarrow k}/2$. The structure of $\rho_{\rm cross}^{ij,(2)}(0)$ explains the reason for $\rho_{\rm cross}(0)$ to be non-zero with more than two spins, as in Fig.~\ref{fig:delayed_nobias_corr_5SMTJ}(f),  and  $\rho_{\rm cross}(0)$ to be zero with only two spins, as in Fig.~\ref{fig:delayed_nobias_corr}(f). When there is no third independent spin $S_{k}$ acting on both $S_{i}$ and $S_{j}$, as when there are only two spins, the long time delay destroys their direct cross-correlation; in contrast, when there are more than two spins, there exists independent $S_{k}$ that provide synchronous spin information to $S_{i}$ and $S_{j}$, indirectly correlating $S_{i}$ and $S_{j}$. Equation~\eqref{eq:main_1st_cross_origin}  shows that for the antisymmetrically coupled spins $\rho_{\rm cross}^{ij,(1)}(0)$ is zero,  but $\rho_{\rm cross}^{ij,(2)}(0)$ can be non-zero if there exist delays according to Eq.~\eqref{eq:main_2nd_cross_origin}. antisymmetrically coupled spins have the delay time dependence $\rho_{\rm cross}^{ij,(1)}(0) = 0$ and $|\rho_{\rm cross}^{ij,(2)}(0)| \sim  1-\exp{(-4 \lambda^{0} t_{d})}$, while symmetrically coupled spins have $|\rho_{\rm cross}^{ij,(1)}(0)| \sim \exp{(-2 \lambda^{0} t_{d})}$ and $|\rho_{\rm cross}^{ij,(2)}(0)| \sim  1+\exp{(-4 \lambda^{0} t_{d})}$. The numerical results shown in Figs.~\ref{fig:corr_peaks}(a) and \ref{fig:corr_entropy_J}(a) agree well with Eqs.~\eqref{eq:main_1st_cross_origin} and \eqref{eq:main_2nd_cross_origin}.

In addition to equal-time cross-correlations, we are also interested in the cross-correlations at $t_{\rm lag} = t_{d}$, corresponding to the cross-correlation functions at the first peak, since they contribute to equal-time cross-correlations. Eq.~\eqref{eq:main_1st_cross_peak} shows that the time delay exponentially suppresses the contribution of $\mathcal{J}_{j \leftarrow i}$ to $\rho_{\rm cross}^{ij,(1)}(t_{d})$ while it does not affect $\mathcal{J}_{i \leftarrow j}$. In the long delay time limit, $\rho_{\rm cross}^{ij,(1)}(t_{d}) \approx \mathcal{J}_{i \leftarrow j} / 2$ survives, making $\rho_{\rm cross}^{ij,(1)}(t_{d})$ dependent on the first order of coupling strength, and $\rho_{\rm cross}^{ij,(2)}(t_{d}) \approx 0$. This analysis is consistent with the numerical results in Fig.~\ref{fig:corr_peaks}(b) and \ref{fig:corr_entropy_J}(c). The autocorrelation function at $t_{\rm lag} = t_d$ has no first order corrections, and both its zeroth order and second order corrections decay with increasing $t_d$ (see Appendix.~\ref{sec:corr_prob_time_deriv}), so that in the long time limit, it depends on coupling strength to the third order. In Fig.~\ref{fig:corr_peaks}(d), $\rho_{\rm auto}(t_{d})$ calculated with up to second order agrees with simulations to all orders and decays indeed exponentially with $t_d$. Fig.~\ref{fig:corr_entropy_J}(d) highlights the higher order behavior of the autocorrelation function. The calculation to second order is zero, shown as the solid line, while the simulated data roughly follow a scaling with the coupling strength to third order, confirming our order counting of $\rho_{\rm auto}(t_{d})$.

The expansion of the probability distribution of joint states from the uniform distribution (see Appendix.~\ref{sec:corr_prob_time_deriv} for details), denoted by $\delta P = P(s_{1}\cdots s_{N}) - 2^{-N}$  to first and second order in the coupling strength is 
\begin{subequations}
    \begin{align}
    & \delta P^{(1)}  = \frac{e^{-2\lambda^{0} t_{d}} }{2^{N}} \sum_{i < j} s_{i}\frac{ \mathcal{J}_{i \leftarrow j} + \mathcal{J}_{j \leftarrow i}}{2}s_{j},\label{eq:main_1st_delta_P} \\
    &\delta P^{(2)} = \frac{1}{2^{N}}\left\{ \frac{1}{4}\sum_{i < j}  s_{i}s_{j}  \sum_{k\neq i, j}  \left[ 2 \mathcal{J}_{i \leftarrow k} \mathcal{J}_{j \leftarrow k} \right.\right. \nonumber \\
    &\left. + (\mathcal{J}_{i \leftarrow k} \mathcal{J}_{k \leftarrow j} + \mathcal{J}_{j \leftarrow k} \mathcal{J}_{k \leftarrow i}) e^{-4 \lambda^{0} t_{d}} \right]  +   \frac{e^{- 4\lambda_{0}t_{d}}}{8} \nonumber \\
    & \left. \times \sum_{|A|=4} \sum_{(i,j,l,m) = \text{Perm}(A)} \mathcal{J}_{i \leftarrow j} \mathcal{J}_{l \leftarrow m} \prod_{k \in A} s_{k} \right\}, \label{eq:main_2nd_delta_P}
    \end{align}
\end{subequations}
where $|A| = 4$ means a set $A=\{i,j,l,m\}$ consisting of four spin indices, $\text{Perm}(A)$ means a permutation of spin indices in $A$, and $\delta P^{(n)}$ is the correction to $\delta P$ to the $n^{\rm th}$ order of coupling strength, with $\delta P^{(0)} = 0$. When delay times are zero and all couplings are symmetric, Eq.~\eqref{eq:main_1st_delta_P} just corresponds to the energy of the spin state, consistent with the first-order expansion of the Boltzmann distribution. 

The expansion of the probability distribution has similarities to that of the cross-correlation function. 
The presence of delay exponentially suppresses the first order correction to $\delta P^{(1)}$ as $\exp{(-2 \lambda^0 t_d)}$, so that in the long delay limit, $\delta P^{(1)}\approx 0$. Regardless of delay time, when all couplings are antisymmetric, $\delta P^{(1)}$ is always zero. In constrast, the second order deviation $\delta P^{(2)}$, survives in the long delay time limit, where $\delta P^{(2)} \approx 2^{-(N+1)} \sum_{i<j}\sum_{k\neq i,j}\mathcal{J}_{i\leftarrow k} \mathcal{J}_{j\leftarrow k}s_{i} s_{j}$. This behavior in $\delta P^{(2)}$ is the same that in the cross-correlation $\rho_{\rm cross}^{ij,(2)}(0)$; both due to the third spin's impact. 

When all couplings are antisymmetric, $\delta P^{(2)} = 0$ in the absence of delay, but in contrast to $\delta P^{(1)}$, $\delta P^{(2)} \neq 0$ as $t_d$ increases, approximately as $\delta P^{(2)} \sim 1-\exp{(-4 \lambda^{0} t_{d})}$. The consequence of non-zero $\delta P$ is reflected in the entropy: $\delta P^{(1)}$ and $\delta P^{(2)}$ give rise to the entropy reduction from maximum value $N \ln 2$ varying as the second and fourth orders of the coupling strength, respectively. For symmetrically coupled spins, increasing the delay time $t_d$ exponentially suppresses the entropy reduction from second order to fourth order; for antisymmetrically coupled spins, the role of delay is different, since in the absence of delay, $\delta P^{(1)} = \delta P^{(2)} = 0$, the entropy reduction is either 0 or even higher than fourth order in the coupling, however, by increasing $t_d$, the entropy reduction grows approximately as $[1-\exp{(-4 \lambda^{0} t_{d})}]^{2}$ and saturates to fourth order in the long delay time limit. The different delay time dependence of entropy reduction is consistent with the curves in Fig.~\ref{fig:delayed_entropy_5SMTJ}(b), and the predicted second-order or fourth-order dependence of entropy reduction on the coupling strength without delay or with long delay is confirmed by Fig.~\ref{fig:corr_entropy_J}(b).

Here, we only consider uniform delay times, all-to-all ferromagnetic or antisymmetric couplings with uniform absolute values. The degrees of freedom regarding the delay times and couplings are much greater than for two spins. For $N$ spins, there are $N(N-1)/2$ symmetric coupling parameters and $N$ biases. Allowing for asymmetric coupling increases the coupling parameters to $N(N-1)$. Allowing for delay times adds $N(N-1)$ additional parameters. We expect there will be more complicated dynamics and steady-state distributions associated with general asymmetrically coupled spins with non-uniform time delays. However, the main conclusions about suppressed non-uniformity of probability distributions and enhanced temporal correlations hold, as illustrated by the numerical results in Appendix~\ref{sec:differnt_patterns} with more complicated couplings and delay times as well as generalized Eq.~\eqref{eq:main_1st_delta_P} of $\delta P ^{(1)} = 2^{-N} \sum_{i < j} (\mathcal{J}_{i \leftarrow j} e^{-2 \lambda^0 t_{i \leftarrow j}} +  \mathcal{J}_{j \leftarrow i} e^{-2 \lambda^0 t_{j \leftarrow i}} ) s_i s_j /2 \approx 0$ with $\lambda^0 \text{min}(t_{i \leftarrow j}, t_{j \leftarrow i}) \gg 1$.

\section{Discussion and Conclusion} \label{sec:conclusion} 

In this work, we investigate generalized Ising models with both asymmetric and time-delayed interactions between the spins. In a coupled two-spin system, antisymmetric couplings produce weak oscillations in both the temporal auto- and cross-correlation functions. However, these oscillations are weaker than those measured in an experiment implementing such coupling. Introducing time delay into the coupling enhances the oscillations to a degree consistent with experimental observations. 

In the long-time limit, interaction delays significantly longer than the intrinsic time scales of the Ising spins lead to nearly uniform probability distributions for all forms of coupling. While this distribution is consistent with weak coupling, the auto- and cross-correlation functions exhibit significant structure inconsistent with that assumption. We theoretically prove that long delays suppress the non-uniformity of probability distribution from first order to second order in the coupling strength for a wide range of models. However, in models that break $\mathbb{Z}_2$ symmetry with bias, the distribution becomes strongly non-uniform and dependent on not just the bias, but also the couplings and delay time. Through numerical simulations of 5-spin systems, we show that our conclusions are applicable to systems with multiple coupled spins. Our work provides critical knowledge in the future design of neuromorphic algorithms, hardware, and helps understand real-world complex networks.

The findings of this work indicate that neuromorphic hardware can still usefully operate in regimes in which hardware imperfections are larger than previously thought prohibitive.
Prior work~\cite{aadit2022massively} has shown that reducing the intrinsic time of each probabilistic bit to be moderately smaller than their communication time ($t_{i\leftarrow j} > 1 /\lambda_{i}^{0}(T) $), greatly reduces the time to solution while maintaining correctness. The key reason such a strategy works can be understood by looking at Fig.~\ref{fig:delayed_no_bias_prob}(e), Eqs.~\eqref{eq:main_1st_delta_P}: when the delay times are uniform, meaning $t_{i \leftarrow j} = t_{j \leftarrow i} = t_d$, and not much larger than the mean dwell time $1/\lambda_{i}^{0}(T)$ of each spin, the probability distribution mimics one with higher temperature or weaker coupling, highlighted by the overall decaying factor $\exp{(-2 \lambda_{i}^{0} t_d)}$ in Eq.~\eqref{eq:main_1st_delta_P}, but the ground states are still distinguishable from other high-energy states, meaning that $R_{\rm log} > 0$, as reflected in Fig.~\ref{fig:delayed_no_bias_prob}(e).

Beyond the upper bound of intrinsic time of Ising spins reported in Ref.~\cite{aadit2022massively}, an often overlooked factor is the uniformity of delay times. The probability distribution of fully antisymmetrically coupled spins is no longer uniform if the delay times are distinct as reflected in Fig.~\ref{fig:delayed_no_bias_prob}(f). More generally, the dependence of the probability distribution to the first order in the couplings is related to $\mathcal{J}_{i \leftarrow j} e^{-2 \lambda^0 t_{i \leftarrow j}} +  \mathcal{J}_{j \leftarrow i} e^{-2 \lambda^0 t_{j \leftarrow i}}$, and each individual interaction $\mathcal{J}_{i \leftarrow j}$ needs to be combined with its own delay time $t_{i \leftarrow j}$ when considering probabilities and correlations. Such dependence of probability distribution on delay times can be utilized as new tuning parameters of probability distributions and temporal correlations.
Another hypothesis from this work is that the long delay times and resulting enhanced temporal correlation suggest  delayed coupled Ising systems have long memories and behave in ways similar to the brain, whose neural activities exhibit various rhythms~\cite{buzsaki2006rhythms}. In addition, learning in the brain, or the nervous system plasticity, is also believed to be mediated by connections with modifiable delay times~\cite{fields2015new}. Design of new neuromorphic hardware based on programmable asymmetry and delay may potentially enable new functionalities unavailable in conventional hardware, for example it would provide network level memory, although concrete demonstrations of such functionalities are beyond the scope of this work and shall be left to future research. 

More generally, the framework and approach developed in this work benefit the modeling and understanding of many real-world complex networks. For example, although different statistical models based on either symmetric or directional couplings have been developed for biological neural networks, the model combining both asymmetry and delay has been less widely explored
\cite{meshulam2025statistical}. Another inspiring similarity exists between the peaks in our model and the strong oscillations of temporal correlations in biological neural networks. Given the hierarchical time scales of different neuronal dynamics and directional propagation of neural signal~\cite{gerstner2002spiking,dayan2005theoretical}, this may suggest  the necessity of combining both asymmetry and delay to understand the rhythms and collective activation of neurons in our brain.

\begin{acknowledgments}
This work was funded by the National Institute of Standards and Technology, National Science Foundation and Agence Nationale de la Recherche. H.Z. acknowledges support from the George Washington University Professional Research Experience Program (GW-PREP) under NIST financial assistance award 70NANB23H019. S.G. and A.M. acknowledge support under NSF Grant No. CCF-CISE-ANR-FET-2121957. S.G. also acknowledges support under the NIST cooperative agreement 70NANB25139-0 under the University of Maryland. A.M. also acknowledges support under the NIST Cooperative Research Agreement Award No. 70NANB14H209 through the University of Maryland. P.T. and U.E. acknowledge support under the ANR ASTROCOMP Project Award No. ANR-25-CE24-5342-01. The authors thank Jabez J. McClelland, William A. Borders, Alexander J. Grutter and Marcelo Davanco for their invaluable comments on the manuscript.  The computational results in this work were made possible by the Blackbird high performance computer cluster on NIST's Gaithersburg campus and supported by NIST's Research Services Office. The authors used OpenAI ChatGPT (GPT-5.6 and 6) as interactive aids to help explore and cross check mathematical derivations. All content, scientific claims, and conclusions have been reviewed and verified by the authors to ensure accuracy and originality, for which the authors take full responsibility.
\end{acknowledgments}

\appendix

\section{Animations of Cross-Correlation Functions With Changing Asymmetric Coupling Strength} \label{sec:animation}
We animate the autocorrelation functions (Supplemental Movie 1) and cross-correlation functions (Supplemental Movie 2) of asymmetrically coupled spins with varying coupling strengths, consistent with experimental settings. $t_{\rm lag}$ is the time difference between two measurements of spin states and $\tau = 1/\lambda_{\rm flip}^{0}$ is the intrinsic characteristic time scale of each spin. Simulations are performed with $\tau = 4.98~\Delta t$, $\Delta E_{i} = 4~k T$, $t_{1 \leftarrow 2} = t_{2 \leftarrow 1} = 100~\Delta t$ or 0. Each data point gives the mean value averaged over $10^{4}$ ensembles and 1000 time steps after reaching steady state. $95~\%$ confidence intervals of correlation functions are obtained from standard statistical analysis but narrower than the width of the line when not visible.

\begin{figure}[!htbp]
\centering
\includegraphics[width=0.95\linewidth]{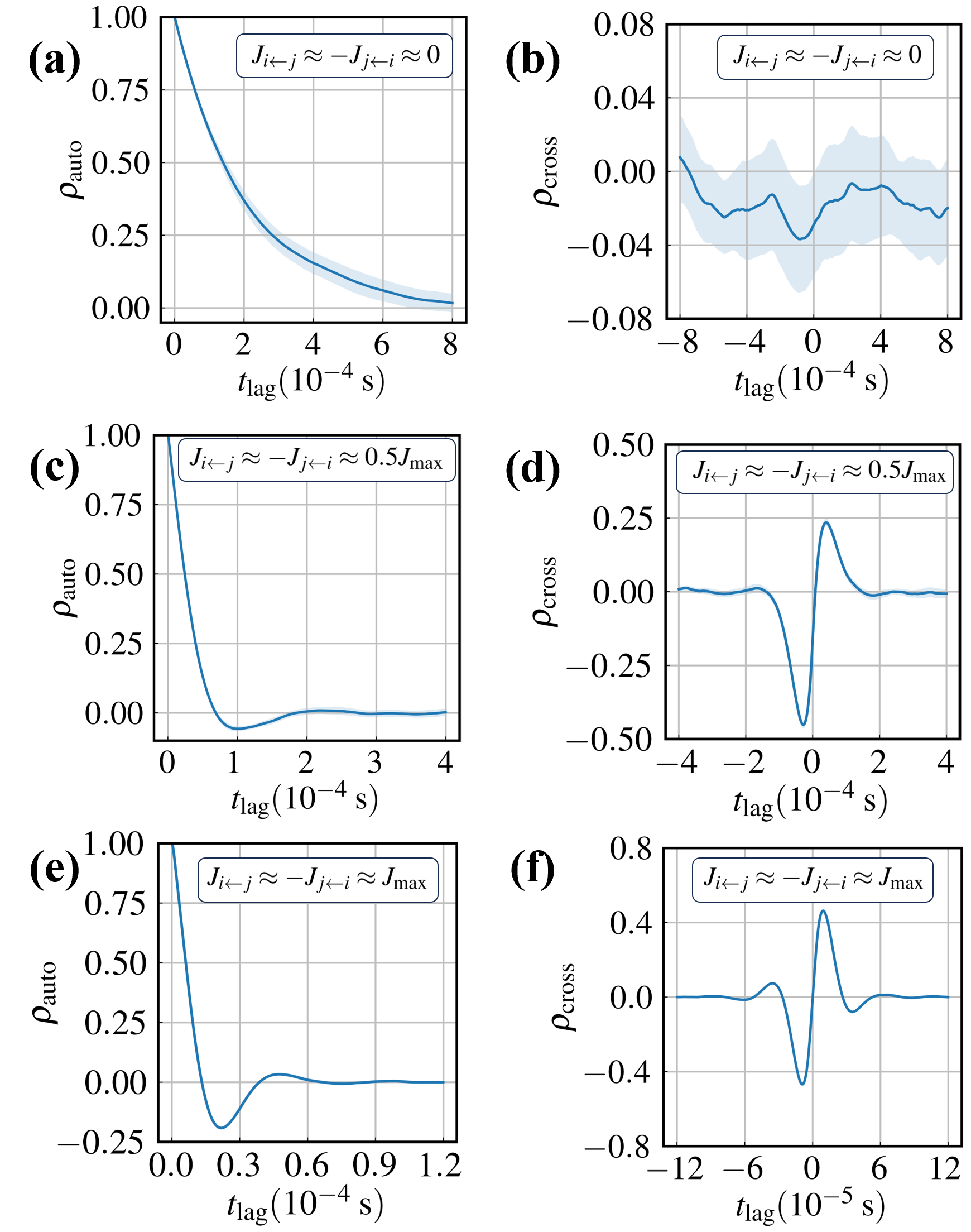}
\caption{Experimental auto- ($\rho_{\rm auto}$) and cross- ($\rho_{\rm cross}$) correlation functions with varying coupling strengths. The coupling between two SMTJs are fully antisymmetric, and the coupling strength is 0 in (a,b), $0.5~J_{\rm max}$ in (c,d), and $J_{\rm max}$ in (e,f). $J_{\rm max}$ is the maximum possible coupling allowed by the circuit. $t_{\rm lag}$ is the time difference between two
measurements of digitized SMTJ voltage states. $95~\%$ confidence intervals of correlation functions obtained from standard statistical analysis are plotted as envelopes.
}
\label{fig:exp_corrs}
\end{figure}

We also plot the experimental auto- and cross-correlation functions in Fig.~\ref{fig:exp_corrs} with varying coupling strength as a comparison. It is clear that with increasing coupling strength, the oscillations are enhanced as more peaks appear. This trend is consistent with what is shown in the animations. One thing to note is that in Fig.~\ref{fig:exp_corrs}(d), the cross correlation is not zero when $t_{\rm lag}=0$ and is not fully antisymmetric around the origin. This is due to the fact that our two SMTJs have slightly different mean dwell times. As a result, the cross correlation function is slightly shifted away from the origin and deviates from an antisymmetric function. Noting that $\langle S_1(t) S_2(t-t_\text{lag} )\rangle=-\langle S_2(t) S_1(t+t_\text{lag} )\rangle$, restores the underlying symmetry of the system.

\section{Cross-Correlation Function and Probability Distribution With Varying Time Delay} \label{sec:corr_prob_time_deriv}

In this appendix, we derive the delay-time-dependence of the correlation functions. To simplify the derivation, we consider the coupled system with uniform quantities of $\lambda^0_{i}(T)=\lambda^0$, $t_{i \leftarrow j } = t_{j \leftarrow i} = t_{d}$, $h_{i} = 0$, and we consider a dimensionless coupling,  $\mathcal{J}_{i \leftarrow j}={J}_{i \leftarrow j}/kT$. The derivations herein justify Eqs.~\eqref{eqs:cross-coors} of the main text: in particular, Eqs.~\eqref{eqs:cross-coors}, which four equations respectively give the first- and second-order corrections to the equal-time cross-correlation $\rho_\text{cross}(0)$ and the first- and second-order corrections to the cross-correlation $\rho_\text{cross}(t_d)$ at the coupling delay $t_d$, are found within the present appendix at Eq.~\eqref{eq:1st-cross-decay}, Eq.~\eqref{eq:2nd_cross_origin}, Eq.~\eqref{eq:1st_cross_peak}, and Eq.~\eqref{eq:2nd_cross_peak}.

\subsection{Uniformity of the marginal distributions}
We begin by proving that the steady-state probability of any single spin's state is $P(s_{i}=\pm 1) = 1/2$ in the absence of bias, since we will use this conclusion throughout the subsequent derivations. When $h_i = 0$, Eq.~\eqref{eq:flipping_rate_delay} is invariant under the action of the $\mathbb{Z}_2$ group on all current and past spin states---that is, flipping all spins (at all times) together yields an equivalent system---leading to $P(s_1,\cdots, s_{N}) = P(-s_{1},\cdots, {-s_{N}})$. Consider then the single-spin steady-state probability $P(s_1)$ as an example; expressing it as a marginalization $P(s_1)=\sum_{s_2,\cdots,s_n} P(s_1,\cdots,s_n)$ of the joint distribution for all spins, we know that
\begin{align}
    P(s_{1}=1) &= \sum_{s_2,\cdots,s_n} P(s_1=1, s_2, \cdots, s_{N}) \nonumber \\
    &= \sum_{s_2,\cdots,s_n} P(s_1=-1, -s_2, \cdots, -s_{N}) \nonumber \\
    &=\sum_{s_2,\cdots,s_n} P(s_1=-1, s_2, \cdots, s_{N}) \nonumber\\
    &= P(s_1 = -1), \label{eq:symmetry-prob}
\end{align}
where the second line is based on our assumption of global $\mathbb{Z}_2$ symmetry, and the third line comes from redefining the dummy indices as $-s_j\mapsto s_j$. The final line necessarily implies that $P(s_{1}=\pm1) = 1/2$. Similarly, $P(s_{i}=\pm1) = 1/2$ for all $i$.

\subsection{The weak-coupling assumption}
Although in the main text---which chiefly concerns only two coupled spins---we treat the exponential form of transition rate $\lambda_{i}$ exactly, we here must abandon such an exact form in the presence of $N$ coupled spins, since the factorized form of  $\lambda_{i}$  in terms of each pairwise spin interaction becomes combinatorially intractable. Additionally, since the N\'eel-Brown model is based on Kramers' theory of rare-event escapes from a potential well~\cite{kramers1940brownian,brown1963thermal}, it assumes the energy change induced by an effective field  $\mu B_{i}S_{i} = \sum_{j\neq i} \mathcal{J}_{i\leftarrow j} S_{j}S_{i}$ is much smaller than the confining potential height $\Delta E_{i}$. Here we maintain that assumption, which enables the approximation of a magnetic moment as a two-state discrete variable, in order to maintain the link between an abstract Ising model and real spintronics-based Ising machines.  

In the spirit of that weak-coupling assumption, consider the case where $\sum_{j\neq i} |\mathcal{J}_{i \leftarrow j}| \ll 1$ for all $i$.
Perturbatively expand Eq.~\eqref{eq:flipping_rate_delay} to  second order in the couplings as
\begin{align} \label{eq:rate_expansion}
     \lambda_{i}(t) & \approx \lambda^{0} \left[  1 - \sum_{j\neq i} \mathcal{J}_{i \leftarrow j} S_{i}(t) S_{j}(t-t_{d})  \right. \nonumber \\
     & \,\left. + \;\frac{1}{2} \sum_{j,k\neq i} \mathcal{J}_{i \leftarrow j} \mathcal{J}_{i \leftarrow k} S_{j}(t-t_{d}) S_{k}(t-t_{d}) \right];
\end{align}
this expansion will be used repeatedly in the coming derivation.  

\subsection{The Fourier basis on spin-space}
The usual view of a joint distribution function $P(s_1,\cdots,s_N)$ is that it assigns a probability to each possible spin configuration; in other words, we think of $P$ as being specified through a set of basis functions $\xi_{\vec\sigma}$, 
\begin{equation}
P(s_1,\cdots,s_N)=\sum_{\sigma_1,\cdots,\sigma_N}P_{\sigma_1,\cdots,\sigma_N}\underset{\xi_{\vec \sigma}(\vec s)}{\underbrace{\left(\prod_{j=1}^N \delta_{\sigma_js_j}\right)}}
\end{equation}
where the $2^N$ possible assignments of $\vec\sigma$ correspond to all possible joint spin configurations; here $\delta_{\sigma s}$ is the Kronecker delta function. The orthonormal basis functions $\xi_{\vec\sigma}$ are akin to the canonical unit (co)vectors: evaluated at any concrete spin state $\vec s$, exactly one of the $\xi_{\vec \sigma}(\vec s)$ will equal one, and the others will be zero. In what follows, however, we will instead expand the joint probability distribution $P(s_1,\cdots,s_N)$ of the system in a \textit{different} set of $s$-dependent basis functions, which we will call $\chi_A(\vec s)$. These $\chi_A$ correspond to a Fourier basis~\cite{donnell2014analysis} for the system.

Consider the marginal distribution $P(s_1)$ over a single spin. This function has a domain of only two points, $s_1=+1$ and $s_1=-1$, and to align with standard notation we temporarily recast these as Boolean variables $b_1=0$ and $b_1=1$, respectively~\footnote{This brief sidestep into the $b$-language, which corresponds to the additive representation of the group $\mathbb{Z}_2$, is more mathematically natural, as the exponentiation in the Fourier transform naturally gives us a map from the Lie algebra (in $b$-representation) to the Lie group (the $s$-representation).}. In other words, $P(s_1)$ is specified (in the $\{\xi_1,\xi_{-1}\}$ basis) by the two numbers $P_0=P(s_1=1)$ and $P_1=P(s_1=-1)$. Thus $P$ can be represented through the (inverse) discrete Fourier transform as
\begin{align}
    P_b &= \frac{1}{\sqrt{2}}\sum_{k=0}^1 \tilde{P}_k \cdot e^{i\pi kb},\\
    \text{where}\quad \tilde P_k&=\frac{1}{\sqrt{2}}\sum_{b=0}^1 P_b \cdot e^{-i\pi kb};
\end{align}
here $\tilde P$ is the discrete Fourier transform of $P$, and serves as a coefficient when expanding $P$ in a plane-wave basis. If the discrete Fourier transform is unfamiliar to the reader, note that $\tilde P$ is just the usual Fourier transform of $P(b)$ if we constrain its probability density to the allowed discrete domain, $P(b)=P_0\delta(b)+P_1\delta(b-1)$. 

For the full joint distribution, each spin index of $P_{b_1b_2\cdots b_N}$ transforms independently, and we have
\begin{equation}
    P_{\vec b}=\frac{1}{2^{N/2}}\sum_{\vec k\in \mathbb{Z}_2^N} \tilde P_{\vec k}\prod_i e^{i\pi k_i b_i}=\frac{1}{2^{N/2}}\sum_{\vec k\in \mathbb{Z}_2^N} \tilde P_{\vec k}e^{i\pi( \vec k \cdot \vec b)},
\end{equation}
where $\vec{k}$ indexes the corners of the unit hypercube. Note that $\tilde{P}_{\vec{k}}$ denotes the Fourier coefficient in the $\mathbb{Z}_2^N$ space and is not necessarily factorizable across the components of $\vec{k}$. Remembering our association between $\vec{b}$ and $\vec{s}$, we have that $\exp(\pm i\pi b_j)=s_j$ and so we could equally well express $P$ and $\tilde P$ as
\begin{align}
    P_{\vec s}&=\sum_{\vec k} \tilde P_{\vec k} \underset{\chi_{\vec k}(\vec s)}{\underbrace{\left(\frac{1}{2^{N/2}}\prod_j s_j^{k_j}\right)}}\\
    \text{and}\quad \tilde P_{\vec k}&=\sum_{\vec s} P_{\vec s}\chi_{\vec k}(\vec s).
\end{align}
Each factor in the product for $\chi_{\vec k}$ carries an $s_j$ corresponding to a nonzero component of $\vec k$. The geometry of the hypercube is of less interest to us that the combinatorics of these interacting spins, and so we hereon drop the vector notation for $\vec k$ and instead indexing over all possible subsets of the spins, writing
\begin{align}
P(s_1,\cdots,s_N)&=\sum_{A\in\mathcal{P}(\{1,\cdots,N\})} \tilde P_A \cdot \chi_A\label{eq:fourier-expanded-dist}\\
\text{where}\quad \chi_A&=\frac{1}{2^{N/2}}\prod_{i\in A}s_i,\label{eq:chi-def}
\end{align}
and where $\mathcal{P}(\{1,\cdots,n\})$ is the power set of $\{1,\cdots,N\}$, that is, its set of all subsets. Just as plane waves $\exp[i\pi (\vec k\cdot\vec x)]$ form a functional basis in continuous space, here the $\chi_A$ form a functional basis (to include the trivial $\chi_{\varnothing } = 2^{-N/2}$) over the discrete spin space. It is straightforward to check that this basis is a complete (\textit{i.e.}  $1=\sum_A |\chi_A(\vec s)|^2$) and orthonormal (\textit{i.e.} $\sum_{\vec s}\chi_A(\vec s) \chi_{A'}(\vec s)=\delta_{AA'}$) basis. As in many physical problems, it will be convenient to compute the properties of $P(s_1,\cdots,s_N)$ in terms of these Fourier basis functions, that is, by examining the dynamics of the coefficients $\tilde P_A$ at each of the $2^N$ possible ``momenta'' $A$ instead of at each of the $2^N$ explicit spin configurations. 

\subsection{Dynamics of the Fourier coefficients}
In that spirit, we now investigate how the statistics of our Fourier coefficients evolve with time. These coefficients, $\tilde P_A=\sum_{\vec s} P_{\vec s}\cdot \chi_A(\vec s)$, are just expectation values of $\chi_A(\vec s)$ over the spin configuration, that is, $\tilde P_A=\langle \chi_A\rangle$. The temporal evolution of $\left<\chi_{A}\right>$ obeys
\begin{equation} \label{eq:dyn_WH_exp}
    \frac{d\left<\chi_{A}\right>}{dt} = -2 \sum_{i \in A} \left<\chi_{A} \lambda_{i}\right>
\end{equation}
since the transition of any spin $s_i$ with $i\in A$ results in $ \chi_{A}\to - \chi_{A}$, and such a transition occurs with probability  $\lambda_{i}\, dt$ within the infinitesimal time window $dt$. Although Eq.~\eqref{eq:dyn_WH_exp} explicitly contains neither time delays $t_d$ nor coupling strengths $\mathcal{J}_{i \leftarrow j}$, these enter into the dynamics through $\lambda_{i}$ according to Eq.~\eqref{eq:rate_expansion}. In steady state, $d \left<\chi_{A}\right> / dt = 0$, and combining Eq.~\eqref{eq:dyn_WH_exp} with \eqref{eq:rate_expansion}, gives to the lowest order in $\mathcal{J}_{i \leftarrow j}$ that
\begin{equation} \label{eq:self_consistent_chi}
      \left<\chi_{A}\right> \approx  \frac{1}{|A|} \sum_{i \in A}\sum_{j\neq i} \mathcal{J}_{i \leftarrow j} \left< \chi_{A}(t) S_{i}(t)S_{j}(t-t_{d}) \right>,
\end{equation}
where $|A|$ means the number of elements in $A$. Equation~\eqref{eq:self_consistent_chi} is a self-consistent equation; to unravel it, we take the common approach of evaluating each higher order correction to the solution in terms of its lower order solutions. Expanding in powers of the characteristic coupling strength $\mathcal{J}$ [\textit{i.e.} assuming that $\mathcal{J}_{i\leftarrow j}/\mathcal{J}=O(1)]$, we have the useful fact that
\begin{align}
\langle S_iS_j\cdots S_k\rangle_0 &= \langle S_i\rangle_0\langle S_j\rangle_0 \cdots \langle S_k \rangle_0,
\end{align}
where $\left<\cdots\right>_{n}$ denotes the ensemble average $\left<\cdots\right>$ truncated to order $O(\mathcal{J}^n)$. We now evaluate $\left<\chi_{A}\right>$ to first order in $\mathcal{J}$, yielding
\begin{align}
    \left< \chi_{A} \right>_{1} &= \frac{1}{|A|} \sum_{i \in A}\sum_{j\neq i} \mathcal{J}_{i \leftarrow j} \left< \chi_{A}(t) S_{i}(t)S_{j}(t-t_{d}) \right>_{0} \nonumber \\
    &= \frac{1}{2^{N/2}|A|} \sum_{i \in A}\sum_{j\neq i} \mathcal{J}_{i \leftarrow j} \left< S_{j}(t-t_{d}) \prod_{\substack{k \in A \\ k\neq i}}S_{k}(t)\right>_{0}.
\end{align}
The expression in bracket has been simplified through the facts that $i\in A$ by construction and that $S_i(t)^2=1$ for Ising spins. Whenever any summation index $j \notin A$,  there is no $S_{j}(t)$ factor in $\prod_{k \in A,k\neq i}S_{k}(t)$. Consequently these $j\notin A$ terms are, $\left< S_{j}(t-t_{d}) \prod_{k \in A, k\neq i}S_{k}(t)\right>_{0} = \left< S_{j}(t-t_{d})\right>_{0} \left<\prod_{k \in A,k\neq i}S_{k}(t)\right>_{0} = 0$ due to the $\mathbb{Z}_{2}$ symmetry (\textit{i.e.} $\langle S_j\rangle=0$  by Eq.~\eqref{eq:symmetry-prob}). It follows that the summed terms in $\left< \chi_{A} \right>_{1}$ are only non-zero for  $j\in A$, giving 
\begin{align} \label{eq:chi_1st_order}
    &\left< \chi_{A} \right>_{1} = \frac{1}{2^{N/2}|A|}\nonumber \\
    &  \times \sum_{\substack{i,j\in A\\i\neq j}} \mathcal{J}_{i \leftarrow j} \left< S_{j}(t-t_{d}) S_{j}(t) \right>_{0} \prod_{\substack{k \in A\\k\notin\{ i,j\}}} \left<S_{k}(t)\right>_{0}.
\end{align}
It turns out that only $A$ with two elements ($|A|=2$) contribute at this order. Note that every nonzero term in the sum above has at least $|A|\geq 2$, because we require that $i\neq j$. But Eq.~\eqref{eq:symmetry-prob} tells us that $\langle S_k\rangle = 0$, and so if $|A|>2$ then the product $\prod_{k\in A, k\notin\{ i,j\}} \left< S_{k}(t)\right>_{0}$ vanishes. Therefore basis functions with $|A| = 2$ have nonzero coefficients at leading order.

And yet, returning to its original definition from Eq.~\eqref{eq:chi-def}, we see that (up to normalization) $\langle \chi_A\rangle$ for $|A|=2$ is just an equal-time cross-correlation function. The cross-correlation function is defined as $\rho^{ij}_{\rm cross}(t,t_{\rm lag}) = \left<  S_{i}(t) S_{j}(t-t_{\rm lag})  \right> / \sigma_{i} \sigma_{j}$, where $\sigma_{i}$ and $\sigma_j$ are the standard deviations of $S_{i}$ and $S_{j}$ in the steady state (in the absence of bias, $\mathbb{Z}_{2}$ symmetry of the system enforces $\left< S_{i}(t) \right> = 0$ as well as $\left< S_{i}^{2}(t) \right> = 1$, so that $\sigma_{i}=1$). A similar form can be written down for the autocorrelation function. We consider here only the steady-state limit where $\rho^{ij}_{\rm cross}(t,t_{\rm lag})$ is independent of $t$, so that $\rho^{ij}_{\rm cross}(t,t_{\rm lag}) = \rho^{ij}_{\rm cross}(t_{\rm lag})$. By choosing $A=\{i,j\}$, Eq.~\eqref{eq:chi_1st_order} tells us the first-order correction to $\rho_{\rm cross}^{ij}(0)=2^{N/2}\langle \chi_A\rangle=\sum_\mu\rho^{ij,(\mu)}_\text{cross}(0) $ is
\begin{align} 
    \rho_{\rm cross}^{ij,(1)}(0) &= \frac{1}{2}\left[ \mathcal{J}_{i \leftarrow j}\rho_{\rm auto}^{jj,(0)}(t_{d}) + \mathcal{J}_{j \leftarrow i}\rho_{\rm auto}^{ii,(0)}(t_{d}) \right]  \label{eq:1st_cross_origin}\\
    & =\frac{e^{-2\lambda^{0} t_{d}}}{2}(\mathcal{J}_{i \leftarrow j} + \mathcal{J}_{j \leftarrow i}),\label{eq:1st-cross-decay}
\end{align}
where $\rho_{\rm auto}^{(n)}$ and $\rho_{\rm cross}^{(n)}$ are the $O(\mathcal{J}^n)$ order corrections to auto- and cross-correlation functions, respectively, and $\rho_{\rm cross}^{ij,(0)} = 0$ and $\rho_{\rm auto}^{ii,(0)}(t_{\rm lag}) = \exp{(-2\lambda^{0} |t_{\rm lag}|)}$.

\subsection{Dynamics of the generalized correlation functions}

To compute second-order corrections to the correlation functions, we need to know $\left< \chi_{A} \right>_{2}$. By performing a similar perturbative expansion for $\left< \chi_{A} \right>_{2}$ as for $\langle \chi_A\rangle_1$, we obtain
\begin{align} %
    &\left< \chi_{A} \right>_{2} = \frac{1}{|A|} \sum_{i \in A}\left[ \sum_{j\neq i} \mathcal{J}_{i \leftarrow j} \left< \chi_{A}(t) S_{i}(t)S_{j}(t-t_{d}) \right>_{1} \right. \nonumber \\
    &\left. - \frac{1}{2} \sum_{j,k \neq i} \mathcal{J}_{i\leftarrow j} \mathcal{J}_{i\leftarrow k} \left< \chi_{A}(t) S_{j}(t - t_{d}) S_{k}(t-t_{d}) \right>_{0}   \right] \nonumber \\
    & = \frac{1}{2^{N/2}|A|} \sum_{\substack{i \in A\\j\neq i}} \mathcal{J}_{i \leftarrow j} \rho^{A_{-i},j,(1)}(t_{d}),\label{eq:gen-corr-def-chi-A}
\end{align}
where the vanishing of the second line occurs because, since $i\neq j$ and $i \neq k$ necessarily, the $\chi_A$ factor always contains an uncanceled $S_i(t)$ term, which factors out of the expectation at zeroth order and gives $\left<S_{i}\right>_{0}=0$ due to $\mathbb{Z}_2$ symmetry. Introducing the notation $A_{-i}=A\setminus \{i\}$ as the set $A$ with index $i$ removed, Eq.~\eqref{eq:gen-corr-def} defines the first-order contribution to the generalized correlation function
\begin{equation}
    \rho^{A_{-i},j}(t,t_{\rm lag}) = \left< S_{j}(t-t_{\rm lag}) \prod_{k \in A_{-i}} S_{k}(t) \right>,\label{eq:gen-corr-def}
\end{equation}
that is, an $|A|$-point correlation function where $S_j$ experiences a relative time lag. The dynamics of this generalized correlation function are the topic of the present section; we will return to $\langle \chi_A\rangle_2$ and use the results we learn here in the next section.

As usual, the steady state limit gives us $\rho^{A_{-i},j}(t,t_{\rm lag}) = \rho^{A_{-i},j}(t_{\rm lag})$.  Differentiating $\rho^{A_{-i},j}(t_{\rm lag})$ with respect to $t_{\rm lag}$ (through the same logic that led to Eq.~\eqref{eq:dyn_WH_exp}) and expanding that derivative up to $O(\mathcal{J})$ gives
\begin{align} 
    &\frac{d \rho^{A_{-i},j,(1)}}{dt_{\rm lag}} = -2 \sum_{l \in A_{-i}} \left<\lambda_{l}(t_{\rm lag})  S_{j}(0) \prod_{k \in A_{-i}} S_{k}(t_{\rm lag})  \right>_{1} \label{eq:1st_dyn_generalized_corr} \\
    &= -2 \lambda_{0} \left[ |A_{-i}| \rho^{A_{-i},j,(1)} - \sum_{l \in A_{-i}} \sum_{m \neq l} \mathcal{J}_{l \leftarrow m} \right. \nonumber \\
    &\left.  \times \left< S_{m}(t_{\rm lag} - t_{d})  S_{j}(0) \prod_{k \in A_{-i,-l}} S_{k}(t_{\rm lag}) \right>_{0}   \right]. \label{eq:dyn_corr_expanded}
\end{align}

We now consider two distinct cases concerning the membership of $j$ in $A$. If $j\notin A$ (and remembering $j\neq i$), then we must have $j=m$ and $A_{-i}=\{l\}$, or else there would once again be lone spin terms that would zero out the product in expectation. In this case, where $A_{-i}$ contains only a single element, $\rho^{A_{-i},j}$ is merely the cross-correlation $\rho_\text{cross}^{lj}$. Then the summation in Eq.~\eqref{eq:dyn_corr_expanded} contains a single term, and the overall dynamics are driven by
\begin{align}
    \frac{d \rho^{A_{-i},j,(1)}}{dt_{\rm lag}} &=-2\lambda_0\rho^{A_{-i},j,(1)}+2\lambda_0 \mathcal{J}_{l\leftarrow j} \rho^{jj,(0)}_\text{auto}(t_\text{lag}-t_d),\label{eq:gen-corr-dyn-j-nin-A}
\end{align}
where the autocorrelation of spin $j$ drives a cross-correlation in spin $l$ via their coupling.

On the other hand, if $j\in A$, $j$ cannot equal to $m$, because in this case, the three-point correlation function $\langle S_{j}(t_{\text{lag}} - t_d) S_{j}(0) S_{j}(t_{\text{lag}}) \rangle$ violates the $\mathbb{Z}_{2}$ symmetry, resulting in $\langle S_{j}(t_{\text{lag}} - t_d) S_{j}(0) S_{j}(t_{\text{lag}}) \rangle = 0$ and zero contributions from the expectation. Then $j \neq m$, and the summation in Eq.~\eqref{eq:dyn_corr_expanded} contains terms of the form $\rho^{jj,(0)}_\text{auto}(t_\text{lag})\langle S_m(t_\text{lag}-t_d)\cdot \prod_{k\in A_{-i,-j,-l}} S_k(t_\text{lag})\rangle_0$. At zeroth order, this expectation has nonzero contributions only where $k=m$ and $A_{-i}=\{j,l,m\}$ with no additional indices in $A_{-i}$ that could serve to zero out the product over $k$ by symmetry. In this case, the summation terms are simply of the form $\rho^{jj,(0)}_\text{auto}(t_\text{lag})\rho^{mm,(0)}_\text{auto}(t_d)$, and the total contribution of the $l$-sum in Eq.~\eqref{eq:dyn_corr_expanded} gives us
\begin{align}
\frac{d \rho^{A_{-i},j,(1)}}{dt_{\rm lag}} &=-6\lambda_0\rho^{A_{-i},j,(1)}+2\lambda_0\rho_\text{auto}^{jj,(0)}(t_\text{lag})\times\nonumber\\&\quad\quad\quad \left[\mathcal{J}_{l\leftarrow m} \rho^{mm,(0)}_\text{auto}(t_d)+ \mathcal{J}_{m\leftarrow l} \rho^{ll,(0)}_\text{auto}(t_d)\right]
\end{align}
where $j$, $m$, and $l$ are the unique elements of $A_{-i}$. 

The initial condition of Eq.~\eqref{eq:1st_dyn_generalized_corr}, which is proportional to Eq.~\eqref{eq:gen-corr-def} evaluated at $t_\text{lag}=0$, necessarily corresponds to an equal-time correlation function among multiple spins. Recall from the discussion following Eq.~\eqref{eq:chi_1st_order} that, at leading order, the equal-time $|A|$-point correlation function, proportional to $\langle \chi_A\rangle_1$, is nonzero only when $|A|=2$. Which spins are involved in Eq.~\eqref{eq:gen-corr-def}? There are two cases: if $j\in A_{-i}$, then the $S_j$ term outside the product will cancel with one inside the product, so at $t_\text{lag}=0$ we have an expression proportional to $\langle \chi_{A_{-i,-j}}\rangle$, and for this to be nonzero we require that $|A_{-i,-j}|=2$, that is, we require that $|A|=4$. On the other hand, if $j\notin A_{-i}$, then Eq.~\eqref{eq:gen-corr-def}
 at $t_\text{lag}=0$ is $\langle \chi_{A_{-i,+j}}\rangle$, so we require $|A_{-i,+j}|=|A|=2$. Taken together, we have the initial condition
\begin{align} \label{eq:initial_condition_gen_rho}
&\frac{\rho^{A_{-i},j,(1)}(0)}{2^{N/2}} =
\begin{cases}
\left< \chi_{A_{-i,+j}} \right>_{1} \delta_{|A|,2} & j \notin A_{-i}\\
\left< \chi_{A_{-i,-j}}\right>_{1} \delta_{|A|, 4} & j \in A_{-i},
\end{cases}
\end{align}
where for all cases other than $|A|\in\{2,4\}$, the initial condition is zero, the dynamics given by Eq.~\eqref{eq:dyn_corr_expanded} becomes $d \rho^{A_{-i},j,(1)} /d t_{\text{lag}} = -2 \lambda_{0} |A_{-i}| \rho^{A_{-i},j,(1)}$, and so indeed such cases' dynamics are trivial at this order.

\subsection{Second-order contributions to cross-correlation at $|A|\in\{2,4\}$}

Since $\rho_{\rm cross}^{ij,(2)}(0)$ corresponds to $\left< \chi_{|A|=2} \right>_{2}$, we first solve Eq.~\eqref{eq:1st_dyn_generalized_corr} with $|A|=2$ and the initial condition given in Eq.~\eqref{eq:1st_cross_origin}. The solution at $t_{\rm lag} = t_d$ is
\begin{equation} \label{eq:1st_cross_peak}
    \rho_{\rm cross}^{ij,(1)}(t_{d}) = \frac{\mathcal{J}_{i \leftarrow j} + \mathcal{J}_{j\leftarrow i}e^{-4\lambda^{0} t_{d}}}{2}.
\end{equation}
Combining Eqs.~\eqref{eq:1st_cross_peak} and \eqref{eq:gen-corr-def-chi-A}, we know that 
\begin{align} \label{eq:2nd_cross_origin}
    \rho_{\rm cross}^{ij,(2)}(0) &= \frac{1}{4}\sum_{k\neq i, j}  \left[ 2 \mathcal{J}_{i \leftarrow k} \mathcal{J}_{j \leftarrow k}  \right. \nonumber \\
    & \left. + (\mathcal{J}_{i \leftarrow k} \mathcal{J}_{k \leftarrow j} + \mathcal{J}_{j \leftarrow k} \mathcal{J}_{k \leftarrow i}) e^{-4 \lambda^{0} t_{d}} \right].
\end{align}
The generalized correlation $\rho^{A_{-i},j,(2)}$ can be obtained by solving a dynamical equation similar to Eq.~\eqref{eq:1st_dyn_generalized_corr}, whose expansion up to order $O(\mathcal{J}^2)$ and initial condition is given by Eq.~\eqref{eq:2nd_cross_origin}. Omitting the tedious algebra, its solution at $t_{\rm lag} = t_d$ is
\begin{align} \label{eq:2nd_cross_peak}
    \rho_{\rm cross}^{ij,(2)}(t_{d}) &= \frac{e^{-2 \lambda^0 t_{d}}}{4}\sum_{k\neq i, j}  \left[ (2 + 4\lambda^0 t_{d}) \mathcal{J}_{i\leftarrow k} \mathcal{J}_{j\leftarrow k} \right. \nonumber \\
    &\left. + \mathcal{J}_{j\leftarrow k} \mathcal{J}_{k\leftarrow i} e^{-4 \lambda^0 t_{d}} + \mathcal{J}_{i\leftarrow k} \mathcal{J}_{k\leftarrow j}  \right].
\end{align}
We can similarly derive $\rho_{\rm auto}^{ii,(1)}(t_{\rm lag})$ and $\rho_{\rm auto}^{ii,(2)}(t_{d})$, which are
\begin{subequations}
\begin{align}
    \rho_{\rm auto}^{ii,(1)}(t_{\rm lag}) &= 0, \label{eq:1st_auto} \\
    \rho_{\rm auto}^{ii,(2)}(t_{d}) &= \frac{e^{-2\lambda^0 t_d} - e^{-6 \lambda^0 t_d}}{4}\sum_{j \neq i} \mathcal{J}_{i\leftarrow j} \mathcal{J}_{j\leftarrow i}. \label{eq:2nd_auto_peak}
\end{align}
\end{subequations}

To calculate the probability distribution, we must also consider terms corresponding to $|A| = 4$ for Eqs.~\eqref{eq:gen-corr-def} and \eqref{eq:1st_dyn_generalized_corr}. To have a nonzero initial condition when $|A|=4$ requires, from Eq.~\eqref{eq:initial_condition_gen_rho}, that $j\in A_{-i}$; otherwise the initial condition is zero. Furthermore, dynamics for $|A|=4$ lack a driving term when $j\notin A_{-i}$, leading to a zero solution at all times. In the nontrivial case that $j \in A_{-i}$, we can solve Eq.~\eqref{eq:1st_dyn_generalized_corr} with the initial condition of \eqref{eq:initial_condition_gen_rho}, and the solution is
\begin{equation}
    \rho^{A_{-i},j,(1)}(t_{\rm lag}) = \frac{e^{-2\lambda_{0}(t_{\rm lag}+t_{d})}}{2} \sum_{l,m \in A_{-i, -j}} \mathcal{J}_{l\leftarrow m}.
\end{equation}
Setting $t_{\rm lag} = t_d$, we have for $|A|=4$,
\begin{align}
    \left< \chi_{|A|=4} \right>_{2} &= \frac{e^{- 4\lambda_{0}t_{d}}}{2^{3+N/2}} \sum_{(i,j,l,m) = \text{Perm}(A)} \mathcal{J}_{i \leftarrow j} \mathcal{J}_{l\leftarrow m},
\end{align}
where $\text{Perm(A)}$ denotes the set, of size $|A|!=24$, of all permutations of $A$. 

\subsection{Deviation from the uniform distribution}
With $\left<\chi_{A}\right>$ in hand, we are ready to calculate the probability distribution. The deviation of a spin state's probability from uniform distribution can be computed by expanding the full distribution in the complete orthogonal basis of the $\chi_A(\bm s)$ functions, as in Eq.~\eqref{eq:fourier-expanded-dist}. At zeroth order, since the distribution is uniform, we must have Fourier coefficient $\tilde P_\varnothing=2^{-N/2}$ in order to give $P(s_1,\cdots,s_N)=2^{-N}$, and all other coefficients $\tilde P_A=\langle\chi_A\rangle$ are equal to zero. Deviations arise when the other Fourier coefficients take on nonzero values; at each $\bm s$, the deviation is
\begin{align} \label{eq:prob_WH_basis}
    \delta P \equiv P(s_{1},\cdots, s_{N}) - \frac{1}{2^{N}} & =\sum_{A \neq \varnothing} \left< \chi_{A}\right> \chi_{A}(\bm s).
\end{align}
Substituting $\left< \chi_{A} \right>_{1}$ and $\left< \chi_{A} \right>_{2}$ into Eq.~\eqref{eq:prob_WH_basis}, we obtain the first and second order corrections to $\delta P$ as
\begin{subequations}
    \begin{align}
    & \delta P^{(1)} = \frac{e^{-2\lambda^{0} t_{d}} }{2^{N}} \sum_{i < j} s_{i}\frac{\mathcal{J}_{i \leftarrow j} + \mathcal{J}_{j \leftarrow i}}{2}s_{j}, \label{eq:prob_1st_order} \\
    &\delta P^{(2)}(s_{1}\cdots s_{N}) = \frac{1}{2^{N}}\Bigg\{ \frac{1}{4}\sum_{i < j}  s_{i}s_{j}  \nonumber \\
    &   
    \times \sum_{k\neq i, j}  \left[ 2 \mathcal{J}_{i \leftarrow k} \mathcal{J}_{j \leftarrow k} + (\mathcal{J}_{i \leftarrow k} \mathcal{J}_{k \leftarrow j} + \mathcal{J}_{j \leftarrow k} \mathcal{J}_{k \leftarrow i}) e^{-4 \lambda^{0} t_{d}} \right]   \nonumber \\
    &  +   \frac{e^{- 4\lambda_{0}t_{d}}}{8}\sum_{|A|=4} \sum_{\substack{(i,j,l,m) \\\in \text{Perm}(A)}} \mathcal{J}_{i \leftarrow j} \mathcal{J}_{l \leftarrow m} \prod_{k \in A} s_{k} \Bigg\}, \label{eq:prob_2nd_order}
    \end{align}
\end{subequations}
and by definition $\delta P^{(0)} = 0$.

Before we conclude this section, we point out the order of coupling dependence of correlation functions, probabilities, and entropy in the long delay time limit ($\lambda^0 t_d \gg 1$). From the above analysis, it is straightforward to obtain the order dependence of $\rho_{\rm cross}^{ij}(0)$, $\rho_{\rm cross}^{ij}(t_d)$, $\rho_{\rm auto}^{ii}(t_d)$ and $\delta P$ as
\begin{subequations}
\begin{align}
     \rho_{\rm cross}^{ij}(0)  &\sim O(\mathcal{J}^{2}), \\
     \rho_{\rm cross}^{ij}(t_d)  &\sim O(\mathcal{J}), \\
     \rho_{\rm auto}^{ii}(t_d)  &\sim O(\mathcal{J}^{3}), \\
     \delta P &\sim O(\mathcal{J}^{2}).
\end{align}   
\end{subequations}

The definition of entropy can be expanded through the same order-by-order expansion as the distribution to give
\begin{align}
    &S_{\rm entropy} = - \sum_{s_{1}\cdots s_{N}} P(s_{1}\cdots s_{N}) \ln P(s_{1}\cdots s_{N}) \nonumber \\
    & \approx N \ln 2 + (N \ln 2-1) \sum_{s_{1}\cdots s_{N}} \delta P - 2^{N-1} \sum_{s_{1}\cdots s_{N}} (\delta P)^{2}. 
\end{align}
Because  $\sum_{s_{1}\cdots s_{N}} \delta P = \sum_{s_{1}\cdots s_{N}} [P(s_{1}\cdots s_{N}) - 2^{-N}] = 1 - 1 = 0$, we conclude that reduction of the entropy  from its maximal (uniform distribution) value is, in the long delay time limit, bounded above by
\begin{equation}
    N \ln 2 - S_{\rm entropy} \sim (\delta P)^{2} \sim O(\mathcal{J}^{4}). 
\end{equation}

\section{Proof of Suppressed Non-uniformity of Distribution for $\mathbb{Z}_n$, $\text{SO}(n)$ and $\text{SU}(n)$ Symmetric Spins With Significant Time Delay} \label{sec:Zn_spin_proof}

We now generalize our analysis of cross-correlation functions and probability distributions from Ising systems to broader classes of spin systems. Before setting up the specific model to be solved, we mention several physically-motivated assumptions that we take throughout this appendix. First, we restrict ourselves to quadratic two-body interactions and the interactions do not involve temporal or spatial derivatives of spins. Second, we suppose that the transition rates between spin states possess a valid power series expansion about the weak-coupling regime (\textit{e.g.} are governed by an exponential of the interaction energy), akin to the main text. Third, we assume that the auto- and cross-correlation functions of the spins obey first-order linear differential equations (\textit{e.g.} no non-linear terms of correlation functions) where the only derivatives are taken with respect to time, and that autocorrelations decay to zero at long times.

We will suppose that each spin $S_{i}$ in our generalized system takes values in a state space $\mathcal{M}$, and that the system is invariant under the (spinwise) action of any $g \in G$, where $G$ is a compact topological group that acts on $\mathcal{M}$. We emphasize that $G$ is the group symmetry over the spin space, and is decoupled from any notion of a spatial lattice on which the spins may or may not reside. We will also assume that $G$ acts transitively on $\mathcal{M}$, that is, any two points in $\mathcal{M}$ are connected by some $g\in G$. For instance, the full group of rotations SO(3) acts transitively on a classical Heisenberg spin, where $\mathcal{M}$ is the 2-sphere; however, the subgroup of rotations about $\hat z$ is \textit{intransitive}, because such rotations cannot connect two points of different polar angle on $\mathcal{M}$. We will return to non-transitive group symmetries later in the appendix. 

Before diving into the analysis, however, we first return to the Ising system analysis of Appendix~\ref{sec:corr_prob_time_deriv}, explaining how a simple representation theory would have arisen from that system, and how it would gave given rise to the basis vectors $\chi_A$ that played a key role in that analysis. Our intent is to offer a conceptual touchstone for those who have not utilized representation theory in detail, but have a broad understanding of its purpose. The expert may skip it, while for those entirely unfamiliar with representation theory, several excellent reviews and textbooks may offer a starting point.

\subsection{Intuition from the Ising case}

We wish to write down the generalized expression for the probability distribution, and to do this we first choose a basis for the functional space within which the probability distribution resides. To build intuition from the Ising case, the symmetry group $\mathbb{Z}_2$ on a single Ising spin is generated by a single operation: the spin-flip function. The regular representation of this group (recall that a representation is a set of matrices whose algebraic relationships faithfully preserve those of the abstract group being represented) is two-dimensional, as there are only two possible spin states, and we could envision it as corresponding to a vector space where $(1,0)$ represents the spin-up state $\left|s_{i} = 1\right>$ and $(0,1)$ represents the spin down state $\left|s_{i} = -1\right>$. Then the spin-flip operation is given by Pauli matrix $\sigma_x$. The eigenvectors of $\sigma_x$ represent the two modes of the system that transform differently under spin flips: the symmetric mode and the antisymmetric mode. These different transformation modes correspond to so-called \textit{irreducible representations}. These $\sigma_x$ eigenvectors [$(1,1)$ and $(1,-1)$] can also be regarded as functions that map the spin-state-vector to a scalar, and offer a functional basis which is indexed by the irreducible representations of the group. In this case these basis functions are the trivial observable (which always maps $\left|s_i\right>\mapsto 1$) and the spin state itself ($\left|s_i\right>\mapsto s_i$). 

In Eq.~\eqref{eq:chi-def}, $\chi_{A}$ served as a basis element for the $N$-spin function space precisely by being a product of these single-spin observables. Recall that a series of single-variable Fourier transforms of the joint probability distribution $P(s_1,\cdots,s_N)$ produced the $\chi_A(\vec s)$ as basis functions. The $i^\text{th}$ factor in the product representation of $\chi_A$ was $s_{i}$ if $i\in A$ but was replaced with the identity ($s_i\rightarrow 1$) if $i\notin A$. We can now see those $\chi_A$ correspond to the irreducible representations: since the full spin space is a tensor product of single-spin spaces, functions on this full space can be expressed through single-spin basis functions in each tensor slot. If each spin has two basis functions $\{|s_i\rangle\mapsto 1,|s_i\rangle\mapsto s_i\}$, one corresponding to each irreducible representation of $\mathbb{Z}_2$, then the full space has basis functions 
\begin{align}
    &\Big\{ |s_1\cdots s_N\rangle\mapsto 1\otimes 1\otimes\cdots \otimes 1, \nonumber\\
     &\hphantom{\Big\{}|s_1\cdots s_N\rangle\mapsto s_1\otimes 1\otimes\cdots \otimes 1, \nonumber\\
  &\hphantom{\Big\{|s_1\cdots s_N\rangle}\;\;\vdots  \nonumber\\
   &\hphantom{\Big\{}|s_1\cdots s_N\rangle\mapsto s_1\otimes s_2\otimes\cdots \otimes s_N\Big\}\\
   =\,&\Big\{ |\vec s\rangle \mapsto 1,|\vec s \rangle \mapsto s_1, \cdots ,|\vec s\rangle \mapsto s_1s_2\cdots s_N\Big\},
\end{align}
which $2^N$  functions are just the set of all $\chi_A(\vec s)$. In Appendix~\ref{sec:corr_prob_time_deriv}, the set $A$ indexed whether or not a spin belonged to the product expression in $\chi_A$, and originally arose from the indices $b_i\in\{0,1\}$ that labeled the symmetric and antisymmetric plane waves in the Fourier expansion. We see now that these $b_i$ simply labeled the irreducible representations, and that expanding a function in the set of basis functions generated by the group representation theory means summing over all possible combinations of spin-wise irreducible representations. In general, irreducible representations may have dimension greater than one, in which case each representation comes with a set of scalar basis functions that transform into each other under the group action. In the Ising case, the spin state can be captured entirely on a real line, so this was not necessary---but for, say, a Heisenberg spin, it is necessarily vectorial states that transform into each other, and so an irreducible representation will come with a basis function for each vector component.

To summarize, the eigenvectors of the Ising spin-flip operator---where this operator generates the entire symmetry group---give an orthonormal basis for the representation theory of that system's $\mathbb{Z}_2$ symmetry group, and their $2^N$ possible tensor products give a basis for the representation of the $N$-spin Ising system. We will now generalize away from Ising spins and back to our generic $\mathcal{M}$-valued spins, where one can always find an orthonormal basis for the function space over $\mathcal{M}$ by examining the eigenfunctions of the group's invariant operators, and the analogs to $\chi_{A}$ (which we will call $\chi_{\bm\alpha,\bm\gamma}$, where $\bm\alpha$ is a vector labeling the spin-wise irreducible representations and $\bm\gamma$ a set of vector component indices $\gamma_i$ associated with each $\alpha_i$)  become products over the spin-wise basis functions.

\subsection{Generalized basis functions}

The existence of orthonormal bases and irreducible representations for general groups of interest here is guaranteed by the Peter-Weyl theorem~\cite{liegroupbook}. To be concrete, let an orthonormal basis function be given by $\phi_{\alpha_{i} \gamma_{i}}[S_{i}(t)]$, where $\alpha_{i}$ labels an irreducible representation of $G$ and $\gamma_{i}$ is the index associated with some vector component in the $\alpha_{i}^\text{th}$ representation (\textit{i.e.} $\phi_{\alpha\gamma}=[\vec\phi_\alpha]_\gamma$). In the Ising case, we found an operator $\sigma_x$ that commuted with all matrix elements in the group---which, trivially, were just $I$ and $\sigma_x$. The eigenvectors of this commuting operator gave us our functional basis. In a more general case, such as if $G=\text{SO}(3)$ and $\mathcal{M}$ is the 2-sphere, then we need to find eigenvectors of an operator that commutes with all rotations, and the natural choice is the Laplacian $\nabla^2$. Its eigenvectors gives us basis elements $\phi_{\alpha \gamma}(\theta,\varphi) \propto e^{i \gamma \varphi}P_{\alpha}^{\gamma}(\cos \theta) $ corresponding to spherical harmonics, with $P_{\alpha}^{\gamma}$ the associated Legendre polynomials and $(\theta, \varphi)$ the spherical angles of spin $S$. 

From now on we remain general, and define $\phi_{0\gamma_{i}}(S_{i}) = 1$ corresponding to the trivial representation that exists for any group. For simplicity, we assume the multiplicity of each non-isomorphic irreducible representation is trivial, so that distinct $\alpha$ correspond to non-isomorphic representations; we discuss nontrivial multiplicities at the end of this section. Under these conditions, Eqs.~\eqref{eq:chi-def} and \eqref{eq:prob_WH_basis}  generalize to~\cite{donnell2014analysis}
\begin{subequations}
    \begin{align}
        \chi_{\bm{\alpha}, \bm{\gamma}}(\bm{s}) &= \prod_{i = 1}^{N} \phi_{\alpha_{i}\gamma_{i}}(s_{i}), \label{eq:group_chi} \\
        P(\bm{s}) &=  P^{(0)} \left[ 1 + \sum_{\bm{\alpha} \neq 0, \bm{\gamma}} \left<  \chi_{\bm{\alpha}, \bm{\gamma}}(\bm{s})\right> \chi_{\bm{\alpha}, \bm{\gamma}}^{*}(\bm{s})   \right], \label{eq:group_prob}
    \end{align}
\end{subequations}
where $\bm{s} = (s_{1},\cdots ,s_{N})$ with similar definitions for $\bm{\alpha}$ and $\bm{\gamma}$, and where $P^{(0)}$ is the uniform probability distribution. As in the Ising case, $\phi_{\alpha_{i}\gamma_{i}}(s_{i})$ represents a mode for which the spin state $s_i$ transforms under a symmetry operation. Comparing Eqs.~\ref{eq:group_prob} with Eqs.~\eqref{eq:chi-def} and \eqref{eq:prob_WH_basis}, these observables $\phi_{\alpha_{i}\gamma_{i}}(s_{i})$ replace the structural role served by the binary spins in our Ising model analysis. In what follows, then, let us intuitively view $\phi_{\alpha_{i}\gamma_{i}}(s_{i})$ as ``generalized'' spins in a non-rigorous sense, keeping in mind that they each in fact correspond to some particular measurement of a spin rather than the full spin state.

\subsection{Cross-correlation functions}

In order to derive the coupled spin dynamics, we first need to determine the uncoupled dynamics of an individual spin. The orthogonality of different irreducible representations, which follows from Schur's Lemma~\cite{liegroupbook}, guarantees that a generalized spin state $\phi_{\alpha\gamma}[S_{i}(t)]$ can only transition to states of the form $\phi_{\alpha\gamma'}[S_{i}(t+dt)]$, with $\alpha$ untouched by the transition. Although not mathematically equivalent, an intuitive understanding about this conservation law is familiar from the case of $\text{SO}(3)$ spins: in the absence of bias, a Heisenberg spin has rotational symmetry, so angular momentum (encoded through the irreducible representation label  $\alpha$), is conserved. Mimicking the autocorrelation function of an Ising spin, we define a generalized, symmetry-invariant autocorrelation function as $\rho_{\rm auto}^{ii, \alpha}(t, t_{\rm lag}) = \sum_{\gamma} \left<\phi_{\alpha\gamma}^{*}[S_{i}(t)] \phi_{\alpha\gamma}[S_{i}(t - t_{\rm lag})] \right>$. In the steady state limit, the dynamics of this generalized autocorrelation function takes the phenomenological form
\begin{align} \label{eq:group_auto_uncoupled_dyn}
    \frac{\partial \rho_{\rm auto}^{ii, \alpha}}{\partial t_{\rm lag}} = -\mathcal{W}^{\alpha} \rho_{\rm auto}^{ii, \alpha},
\end{align}
where $\text{Re}[\mathcal{W}^{\alpha} ]> 0$ is a complex constant corresponding to the $\alpha^\text{th}$ representation, assumed to be uniform among all spins. When $t_{\rm lag} \to \infty$, the solution implies $\rho_{\rm auto}^{ii, \alpha} \to 0$. The physical justification of this fact is that, when there is no time-delayed interaction, each individual spin follows a strict Markov process and thus spin does not have a memory of its long past history. 

Our next goal is to perturbatively describe the coupled dynamics of many spins. Suppose we have pairwise interactions, and that the interaction of each spin pair has the symmetry-allowed form 
\begin{align} \label{eq:general_interaction_group}
    V_{i\leftarrow j} = \sum_{\alpha} \mathcal{J}_{i \leftarrow j}^{\alpha} \sum_{\gamma} \phi_{\alpha \gamma}^{*}[S_{i}(t)] \phi_{\alpha \gamma}[S_{j}(t - t_{i\leftarrow j})],
\end{align}
where unitless $\mathcal{J}_{i \leftarrow j}^{\alpha} = J_{i \leftarrow j}^{\alpha} / kT$ represents the coupling between two spins in $\alpha^\text{th}$ representation. The structure of Eq. \eqref{eq:general_interaction_group} is justified in general by two facts: the orthogonality of different irreducible representations prevents any coupling between $\alpha$ and $\alpha'\neq \alpha$, and invariance under $G$ gives us the inner product form $\vec \phi^*_\alpha\cdot \vec \phi_\alpha$, so that applying a group operation (which is always represented as a unitary matrix) to all spins in the system leaves every term in $V_{i\leftarrow j}$ unchanged. Defining a generalized, symmetry-invariant cross-correlation function $\rho_{\rm cross}^{ij, \alpha}(t, t_{\rm lag}) = \sum_{\gamma} \left<\phi_{\alpha\gamma}^{*}[S_{i}(t)] \phi_{\alpha\gamma}[S_{j}(t - t_{\rm lag})] \right>$, similar to Eqs.~\eqref{eq:dyn_WH_exp} and its related perturbative expansions, we can phenomenologically express the dynamics of $\rho_{\rm cross}^{ij, \alpha}(t, 0)$ to the first order of coupling strength as
\begin{align} \label{eq:general_cross_dyn}
   & \frac{\partial \rho_{\rm cross}^{ij, \alpha}(t, 0)}{\partial t} \approx -2 \text{Re}[\mathcal{W}^{\alpha}] \rho_{\rm cross}^{ij, \alpha} + \mathcal{B}^{\alpha} \sum_{k \neq i,\alpha',\gamma,,\gamma'} \mathcal{J}_{i \leftarrow k}^{\alpha'}    \nonumber \\
   &  \times \left< \phi_{\alpha\gamma}^{*}[S_{i}(t)]  \phi_{\alpha\gamma}[S_{j}(t)]  \phi_{\alpha' \gamma'}[S_{i}(t)] \phi_{\alpha' \gamma'}^{*}[S_{k}(t - t_{i\leftarrow k})] \right> \nonumber \\
   & + (i \leftrightarrow j)^{*},
\end{align}
where $\mathcal{B}^{\alpha}$ is a phenomenological constant. We use Eq.~\eqref{eq:general_cross_dyn} to evaluate the first order correction of the cross-correlation function, $\rho_{\rm cross}^{ij, \alpha, (1)}(t, 0)$, which requires the  calculation of $\left< \phi_{\alpha\gamma}^{*}[S_{i}(t)]  \phi_{\alpha\gamma}[S_{j}(t)]  \phi_{\alpha' \gamma'}[S_{i}(t)] \phi_{\alpha' \gamma'}^{*}[S_{k}(t - t_{i\leftarrow k})] \right>_{0}$ with $\left<\cdots\right>_{0}$ denoting an ensemble average in the absence of couplings. Since, without coupling, $S_{i}$ is uncorrelated with $S_{j}$ and $S_{k}$, we can factor the product of expectations and, utilizing the normalization conditions of our basis functions,
\begin{align}
    &\left< \phi_{\alpha\gamma}^{*}[S_{i}(t)]  \phi_{\alpha\gamma}[S_{j}(t)]  \phi_{\alpha' \gamma'}[S_{i}(t)] \phi_{\alpha' \gamma'}^{*}[S_{k}(t - t_{i\leftarrow k})] \right>_{0} \nonumber \\
    &= \int_{\mathcal{M}}\left(\phi_{\alpha\gamma}^{*}[S_{i}(t)] \phi_{\alpha' \gamma'}[S_{i}(t)] P^{(0)}(S) \right)dS \nonumber \\
    &\times \left<   \phi_{\alpha\gamma}[S_{j}(t)]  \phi_{\alpha' \gamma'}^{*}[S_{k}(t - t_{i\leftarrow k})] \right>_{0}\label{eq:haar-integral} \\
    & = \delta_{\alpha, \alpha'} \delta_{\gamma, \gamma'} \delta_{j,k} \left< \phi_{\alpha\gamma}[S_{j}(t)]  \phi_{\alpha' \gamma'}^{*}[S_{k}(t - t_{i\leftarrow k})] \right>_{0} 
\end{align}
where $\delta$ is the Kronecker-delta function, and $dS=d\mu(S)$ denotes the %
measure over the state space $\mathcal{M}$ (\textit{e.g.} $dS = \sin\theta \,d\theta\, d\varphi$ for the 2-sphere). By requiring $\partial \rho_{\rm cross}^{ij, \alpha,(1)}(t, 0)/\partial t = 0$ in the steady state limit, we arrive at
\begin{align} \label{eq:group_1st_cross}
    &\rho_{\rm cross}^{ij, \alpha,(1)}(t, 0) = \frac{1}{2 \text{Re}[\mathcal{W}^{\alpha}]} \nonumber \\
    & \times  \left[ \mathcal{B}^{\alpha}\mathcal{J}_{i \leftarrow j}^{\alpha} \rho_{\rm auto}^{jj, \alpha, (0)*}(t, t_{i \leftarrow j}) + \mathcal{B}^{\alpha*}\mathcal{J}_{j \leftarrow i}^{\alpha*} \rho_{\rm auto}^{ii, \alpha, (0)}(t, t_{j \leftarrow i}) \right].
\end{align}
We can see that when delay times are sufficiently long (\textit{i.e.} as $\text{min}(t_{i \leftarrow j})\to \infty$), each autocorrelation function approaches zero, as does $\rho_{\rm cross}^{ij, \alpha,(1)}(t, 0)$. Taken together, this means that $\rho_{\rm cross}^{ij, \alpha}(t, 0)=O(\mathcal{J}^2)$ has at most a second-order dependence on the coupling strength.

\subsection{Probability distributions}

Although we have calculated the generalized cross-correlation functions, using Eq.~\eqref{eq:group_prob} to calculate a probability distribution is not straightforward, especially if $G$ is a non-Abelian group. To bridge the gap between $\rho_{\rm cross}^{ij, \alpha}(t, 0)$ and $P(\bm{s})$, we first consider possible non-zero contributions from $\left< \chi_{\bm{\alpha}, \bm{\gamma}}(\bm{s}) \right>$ where there are only two non-zero $\alpha_{i}$ and $\alpha_{j}$ such that $\left< \chi_{\bm{\alpha}, \bm{\gamma}}(\bm{s}) \right> = \left< \phi_{\alpha_{i},\gamma_{i}}  \phi_{\alpha_{j},\gamma_{j}}\right>$ ($s_{i}$ and $s_{j}$ now omitted for brevity). We use the vector and matrix notations $\bm{\phi}^{\alpha_{i}}$ and $\bm{Q}^{\alpha_{i}\alpha_{j}}$ to represent $\phi_{\alpha_{i},\gamma_{i}}$ and $\left< \phi_{\alpha_{i},\gamma_{i}}  \phi_{\alpha_{j},\gamma_{j}}\right>$. Suppose we have some group element $g\in G$ whose matrix form within the $\alpha^\text{th}$ representation is $\bm{D}^{\alpha}(g)$. Under the transformation of $g$, $\bm{Q}^{\alpha_{i}\alpha_{j}}$ is mapped to $\bm{D}^{\alpha_{i}}(g) \bm{Q}^{\alpha_{i}\alpha_{j}}[\bm{D}^{\alpha_{j}}(g)]^T$ (where $T$ denotes matrix transpose); symmetry enforces that $\bm{Q}^{\alpha_{i}\alpha_{j}} = \bm{D}^{\alpha_{i}}(g) \bm{Q}^{\alpha_{i}\alpha_{j}}[\bm{D}^{\alpha_{j}}(g)]^T$. Since our irreducible representations (being of compact topological groups) are unitary, we have $\bm{Q}^{\alpha_{i}\alpha_{j}}[\bm{D}^{\alpha_{j}}(g)]^* = \bm{D}^{\alpha_{i}}(g) \bm{Q}^{\alpha_{i}\alpha_{j}}$, which defines $\bm{Q}^{\alpha_{i}\alpha_{j}}$ as an intertwining operator connecting two representations: $\alpha_{i}$, and the conjugate representation of $\alpha_{j}$. 

Applying Schur's lemma to this intertwining relation, $\bm{Q}^{\alpha_{i}\alpha_{j}}$ must be either $0$ or else an isomorphism~\cite{liegroupbook}. If $\bm{Q}^{\alpha_{i}\alpha_{j}}$ is zero, then its corresponding $\left< \chi_{\bm{\alpha}, \bm{\gamma}}(\bm{s}) \right>$ does not contribute to the probability distribution. Assume then that the $\alpha_{i}$ representation is isomorphic to the conjugate representation of $\alpha_{j}$. It follows that, by properly choosing $\alpha_{i}$ and $\alpha_{j}$, $\left< \phi_{\alpha_{i},\gamma_{i}}  \phi_{\alpha_{j},\gamma_{j}}\right>$ becomes $\left< \phi_{\alpha,\gamma_{i}}^{*}  \phi_{\alpha,\gamma_{j}}\right>$ with $\alpha = \alpha_{i} = \alpha_{j}$, denoted by $\bm{Q}^{\alpha}$. The invariance condition now becomes $\bm{Q}^{\alpha}\bm{D}^{\alpha}(g) = \bm{D}^{\alpha}(g)\bm{Q}^{\alpha}$. We can see that $\bm{Q}^{\alpha}$ is not only an intertwining operator but also a mapping of the representation $\alpha$ into itself. Consequently, Schur's Lemma tells us that $\bm{Q}^{\alpha}$, which is by construction the cross-correlation matrix, is proportional to an identity matrix $\bm{I}$,  $\left< \phi_{\alpha,\gamma_{i}}^{*}  \phi_{\alpha,\gamma_{j}}\right> \sim \delta_{\gamma_{i}, \gamma_{j}} \rho_{\rm cross}^{ij, \alpha}(t, 0)$, which we had just above shown to be $O(\mathcal{J}^2)$. In the long delay limit, then, the first order correction to $\left< \phi_{\alpha,\gamma_{i}}^{*}  \phi_{\alpha,\gamma_{j}}\right>$ is zero. 

As we have exclusively considered the two non-zero $\alpha_{i}$ and $\alpha_{j}$, we now show that it is the only relevant case at leading order---that is, that $\left< \chi_{\bm{\alpha}, \bm{\gamma}}(\bm{s}) \right>$ has vanishing first-order correction if there are more than two non-zero $\alpha$ indices. We use three non-zero indices, $\alpha_{i}$, $\alpha_{j}$ and $\alpha_{l}$ as an example; cases with more non-zero indices follow the same spirit. The first-order correction in question is $\left< \chi_{\bm{\alpha}, \bm{\gamma}}(\bm{s}) \right> = \left< \phi_{\alpha_{i},\gamma_{i}}  \phi_{\alpha_{j},\gamma_{j}} \phi_{\alpha_{l},\gamma_{l}}\right>$. Similar to Eq.~\eqref{eq:general_cross_dyn}, we can differentiate it with respect to time $t$, and the first-order terms are proportional to $\left<\phi_{\alpha\gamma}(S_{l}) \phi_{\alpha\gamma}^{*}(S_{i})  \phi_{\alpha\gamma}(S_{j})  \phi_{\alpha' \gamma'}(S_{i}) \phi_{\alpha' \gamma'}^{*}(S_{k}) \right>_{0}$. Since $i\neq j, k, l$, we can separate $\left< \phi_{\alpha\gamma}^{*}(S_{i}) \phi_{\alpha' \gamma'}(S_{i})\right>_{0}$ from the product, and the remaining part is $\left<\phi_{\alpha\gamma}(S_{l})  \phi_{\alpha\gamma}(S_{j}) \phi_{\alpha' \gamma'}^{*}(S_{k}) \right>_{0}$ with 
$l \neq j$. Whether $k = l$, $k = j$, or $k \notin \{ j,l\}$, the fact remains that we can always find one of these three spins whose $\phi_{\alpha\gamma}$ function can be factored in expectation from the remaining ones, \textit{e.g.} $\left<\phi_{\alpha\gamma}(S_{l})  \phi_{\alpha\gamma}(S_{j}) \phi_{\alpha' \gamma'}^{*}(S_{k}) \right>_{0}=\left<\phi_{\alpha\gamma}(S_{l})  \phi_{\alpha\gamma}(S_{j})\right>\left< \phi_{\alpha' \gamma'}^{*}(S_{k}) \right>_{0}$ in the case $k\neq\{j,l\}$ case.  But $\left< \phi_{\alpha\gamma} \right>_{0}$ is zero due to the group symmetry, so all first-order terms vanish, and the total first-order correction to $\left< \phi_{\alpha_{i},\gamma_{i}}  \phi_{\alpha_{j},\gamma_{j}} \phi_{\alpha_{l},\gamma_{l}}\right>$ is zero. Since we have proven that first order corrections to all possible $\left<  \chi_{\bm{\alpha}, \bm{\gamma}}(\bm{s})\right>$ are zero, and according to Eq.~\eqref{eq:group_prob} neither are first order corrections to the deviation from a uniform distribution, any  such deviation in the long delay time limit has at most $O(\mathcal{J}^{2})$ dependence on the coupling strength.

\subsection{Conjectured extensions to degenerate and symmetry-broken systems}

We now remark on two closely related assumptions we used throughout Appendix~\ref{sec:Zn_spin_proof}: first, that the multiplicity of each irreducible representation is trivial; second, that the symmetry group $G$ acts transitively on the state space. We offer the following discussions as informed hypotheses about possible application of our theory to experimentally relevant systems which usually lack the fully symmetry of their space, either by physical nonidealities or, in many cases, by design. However, we emphasize that these discussions are intended to serve as reasoned conjectures; we do not rigorously prove what follows.

\subsection{Multiplicity of irreducible representations}

First, we address the assumption that the multiplicity of each irreducible representation is trivial, which allowed each $\alpha$ index to correspond to a unique irreducible representation. For the case of multiplicity greater than one, we may combine isomorphic irreducible representations into a single reducible representation that is \textit{isotypic} (\textit{i.e.} consists only of irreducible representations that are degenerate to each other) through a direct sum, redefining $\alpha$ and $\gamma$ as indices labeling each isotypic representation. Under this scheme, non-degenerate irreducible representations effectively retain their original labels and properties. 

We conjecture that the consequences of degeneracy are conceptually minor, but involve extra bookkeeping. First, since the group symmetry does not forbid couplings within an isotypic representation, Eqs.~\eqref{eq:group_auto_uncoupled_dyn} is promoted to a matrix equation within each isotypic representation for some constant invertible matrix $\bm{\mathcal{W}}^{\alpha;\mu\nu}$ (where $\mu,\nu$ index within the isotypic representation $\alpha$) whose eigenvalues have strictly positive real parts that couples degenerate irreducible representations. The sum over $\alpha$ in Eq.~\eqref{eq:general_interaction_group} is promoted to triple sum over $\alpha$, $\mu$, and $\nu$, such that $\mu$ and $\nu$ internally index isotypic representation $\alpha$. 

Eq.~\eqref{eq:group_1st_cross} is also generalized to a matrix equation, but its resulting cross-correlation matrix still vanishes at the first order of coupling strength. The reason is that, by lifting our previous reasoning to include the possibility of degeneracy, Schur's lemma requires that $\bm{Q}$ be a block matrix where blocks are nonzero only for $\alpha=\alpha'$ (\textit{i.e.} within an isotypic representation), and the block $\bm{Q}^{\alpha\alpha;\mu\nu}$ is either zero or proportional to an identity matrix. Consider a block of $\bm{Q}$ that corresponds to a single isotypic representation. Since the degenerate irreducible representations of that isotypic block are necessarily equal in dimension, this block of $\bm{Q}$ has a tensor decomposition of the form $\bm{q}\otimes \bm I$, where $\bm{q}$ is simply the cross-correlation matrix among these degenerate representations. But then $\bm{q}=O(\mathcal{J}^2)$, and so in fact $\bm{Q}=O(\mathcal{J}^2)$ as well.  

These considerations suggest that the cross-correlation matrix has vanishing first-order dependence in the long delay time limit, even if the multiplicity is nontrivial. 

\subsection{Degraded symmetries and non-transitivity}

Second, we have assumed that $G$ acts transitively on $\mathcal{M}$. Suppose now that this was not the case, and take as a concrete example of interest a Heisenberg model spin ($\mathcal{M}\cong S^2$) that possesses an easy-axis anisotropy, which becomes relevant when implementing Ising machines with easy-axis magnets. This breaks the full SO(3) symmetry; the new symmetry group is $\text{O}(2)=\text{SO(2)}\ltimes \mathbb{Z}_2 $, which consists of rotations about the easy-axis but, \textit{crucially}, also includes mirror symmetry under $\hat z\leftrightarrow -\hat z$. By properly taking care of emerging multiplicity of irreducible representations of the new group, one is still free to express the cross-correlation functions in a functional basis corresponding to this symmetry group, and the integral from Eq.~\eqref{eq:haar-integral} over the group measure still reduces as prescribed through the orthonormality of those basis functions, resulting in the first-order cross-correlation functions decaying to zero at long time delays through the decay of the single-spin autocorrelations [Eq.~\eqref{eq:group_1st_cross}]. Without working through the full arithmetic here, one can roughly envision that the residual mirror symmetry is essential to maintain the $\langle S_z(t)\rangle=0$ property that seen to be crucial to the derivations of Appendix~\ref{sec:corr_prob_time_deriv}. 

Because the first-order correlations still vanish in the long-delay limit, we have that the distribution factors into its single-spin marginals, $P(s_1,\cdots,s_N)=\prod_j P(s_j)+O(\mathcal{J}^2)$, up to corrections that appear only at second order. Previously, we had that these single-spin marginal distributions were uniformly distributed (\textit{e.g.} in Eq.~\eqref{eq:symmetry-prob} for the Ising case), which is equivalent to the requirement that these distributions are $G$-symmetric. This generalized requirement translates to the non-transitive case; since correlations are suppressed at leading order, the joint distribution must factor into a product of $G$-symmetric marginals, up to $O(\mathcal{J}^2)$. These marginals are straightforward to compute; though they may not be identical to each other (if for instance the easy-axis anisotropy differs among spins in the system), they will be uniquely determined by each spin's Hamiltonian and their product gives the joint distribution up to first-order in $\mathcal{J}$.

\subsection{Further symmetry reduction}

In the previous section, we focused on the case of an easy-axis Heisenberg spin, where the reduced symmetry group still enforced a vanishing mean for all spin components. Now consider the application of an external magnetic field along the $\hat z$-axis. The symmetry is reduced from $\text{SO}(3)$ to the planar rotation group $K = \text{SO}(2)$, breaking the mirror symmetry $\hat z \leftrightarrow -\hat z$. 

To understand how delayed interactions operate in such a system, we can decompose the spin observables according to the irreducible representations of the surviving subgroup $K$, labeled by their azimuthal charge $m \in \mathbb{Z}$. 

First consider the symmetry-broken sector. The longitudinal observable $S_z$ transforms under the trivial representation ($m = 0$) of $\text{SO}(2)$. Because it is unconstrained by any residual parity or rotation, its uncoupled expectation value is generally non-zero ($\langle S_z \rangle_0 \neq 0$). Consequently, the uncoupled temporal autocorrelation does not decay to zero at long delays, but instead approaches $\langle S_z \rangle_0^2$. This delayed memory acts as an effective static field that sustains first-order cross-correlations, matching our observations for biased Ising networks in the main text.
 
In the surviving, symmetric sector, the transverse observables $S^\pm = S_x \pm i S_y$ transform under non-trivial irreducible representations ($m = \pm 1$) of $\text{SO}(2)$. Rotational invariance about $\hat z$ strictly forces their uncoupled expectation values to vanish ($\langle S^\pm \rangle_0 = 0$). As a result, the transverse uncoupled autocorrelations decay to zero, and the first-order cross-correlations between transverse components remain exponentially suppressed by interaction delays.

Because the longitudinal ($m = 0$) and transverse ($m = \pm 1$) sectors belong to non-isomorphic representations of the surviving symmetry group $K$, Schur's lemma guarantees that the equal-time cross-correlation matrix $\bm{Q}$ contains no off-diagonal blocks connecting them. Thus, leading-order cross-correlations between the symmetry-broken and symmetry-preserved sectors vanish identically.

These considerations suggest that even when a physical system experiences significant symmetry breaking, time delay continues to suppress cross-correlations within any sector that retains a non-trivial symmetry, as well as between that sector and the rest of the network. We leave a full and rigorous verification of these branched-symmetry dynamics to future work.

\section{Correlation function and entropy with more complicated interactions and delay times} \label{sec:differnt_patterns}
\begin{figure}
\centering
\includegraphics[width=1\linewidth]{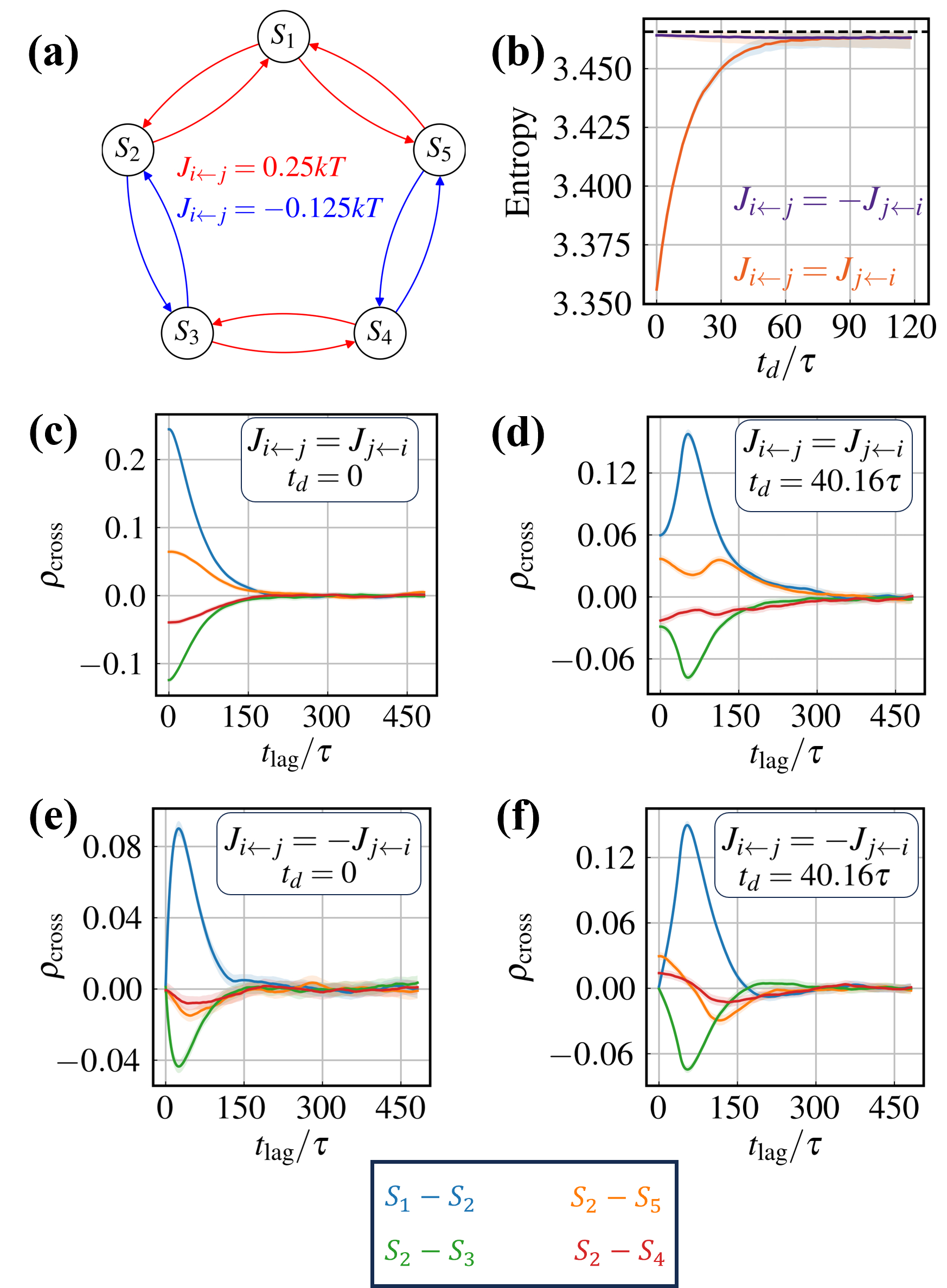}
\caption{Entropy and cross-correlation functions for mixed interactions in a ring geometry. (a) The coupling pattern is shown for the case with symmetric interactions. The coupling for the case with antisymmetric interactions are obtained by reversing the sign of one of the interactions between each pair of neurons, i.1. $J_{j\leftarrow i} = - J_{i\leftarrow j}$. Delay times are uniform, $t_{i\leftarrow j} = t_{j \leftarrow i} = t_{d}$. (b) Dependence of entropy of five symmetrically ($J_{i \leftarrow j} = J_{j \leftarrow i}$) and antisymmetrically ($J_{i \leftarrow j} = -J_{j \leftarrow i}$) coupled spins on the delay time. The dashed line indicates the maximum entropy, $5\ln2$. Cross-correlation functions are shown with symmetric (c,d) or antisymmetric (e,f) interactions, with non-zero delay time (d,f) or without delay (c,e). Different colors correspond to cross-correlation functions of different spin pairs, as shown in the legend at the bottom of the figure. $t_{\rm lag}$ is the time difference between two measurements of spin states. $\tau = 1/\lambda_{\rm flip}^{0}$ is the intrinsic characteristic time scale of each spin in (b-f). Simulations are performed with $\tau = 4.98 \Delta t$, $\Delta E_{i} = 4k T$, $h_i = 0$. Each data point in (b-f) gives the mean value averaged over $10^{4}$ ensembles and $5000~\Delta t$ after reaching steady state. $95~\%$ confidence intervals obtained from standard statistical analysis are plotted as shaded regions.
}
\label{fig:corr_entropy_graph_1}
\end{figure}

\begin{figure}
\centering
\includegraphics[width=1\linewidth]{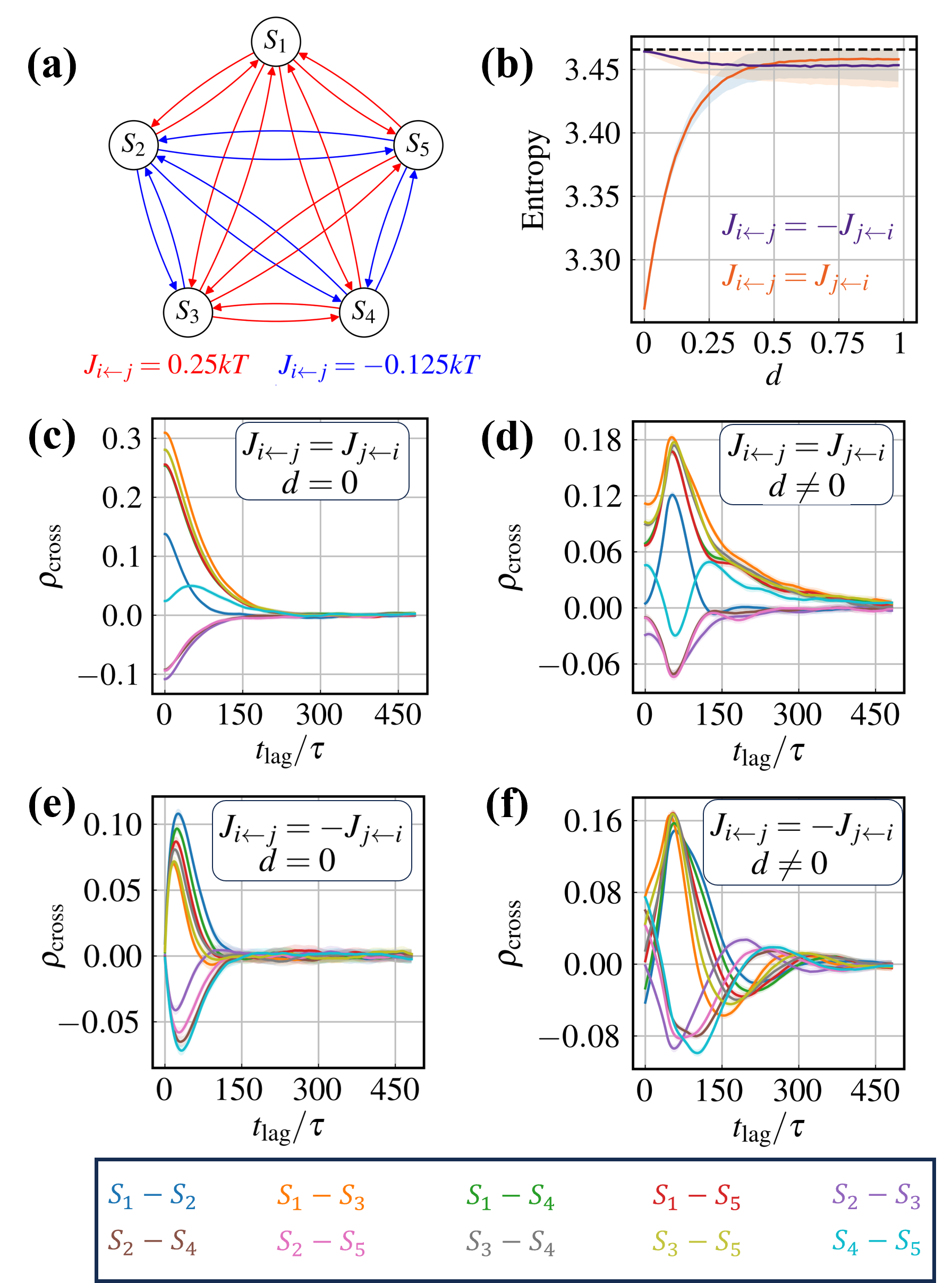}
\caption{Entropy and cross-correlation functions for mixed interactions and delay times in an all-to-all-connected geometry. (a) The coupling pattern is shown for the case with symmetric interactions. The coupling for the case with antisymmetric interactions are obtained by reversing the sign of one of the interactions between each pair of neurons, i.1. $J_{j\leftarrow i} = - J_{i\leftarrow j}$. The matrix of delay times is $t_{i \leftarrow j} = d[\bm{t}^{\rm delay}]_{ij}$ where $d$ is a scaling factor and $\bm{t}^{\rm delay}$ is given in Eq.~\eqref{eq:d_matrix_graph_2}. (b) Dependence of entropy of five symmetrically ($J_{i \leftarrow j} = J_{j \leftarrow i}$) and antisymmetrically ($J_{i \leftarrow j} = -J_{j \leftarrow i}$) coupled spins on the delays as parameterized by the scaling factor $d$. The dashed line indicates the maximum entropy, $5\ln2$. Cross-correlation functions are shown with symmetric (c,d) or antisymmetric (e,f) interactions, with finite delay times ($d=1/3$) (d,f) or without delay ($d=0$) (c,e). Different colors correspond to cross-correlation functions of different spin pairs, as shown in the legend at the bottom of the figure. $t_{\rm lag}$ is the time difference between two measurements of spin states. $\tau = 1/\lambda_{\rm flip}^{0}$ is the intrinsic characteristic time scale of each spin in (b-f). Simulations are performed with $\tau = 4.98 \Delta t$, $\Delta E_{i} = 4k T$, $h_i = 0$. Each data point in (b-f) gives the mean value averaged over $10^{4}$ ensembles and $5000~\Delta t$ after reaching steady state. $95~\%$ confidence intervals obtained from standard statistical analysis are plotted as shaded regions.
}
\label{fig:corr_entropy_graph_2}
\end{figure}

In the main text, we consider the entropy and correlation functions in a five-coupled spin system with uniform interaction strengths and delay times. To check whether our main conclusion is still valid for more general interactions and delay times, we calculate the entropy and cross-correlation function with two different sets of $J_{i\leftarrow j}$ and $t_{i \leftarrow j}$, and show the results for a ring geometry in Fig.~\ref{fig:corr_entropy_graph_1} and and for a fully-connected set Fig.\ref{fig:corr_entropy_graph_2}. In Fig.~\ref{fig:corr_entropy_graph_1} the delay times are all the same but in Fig.~\ref{fig:corr_entropy_graph_2}, they are non-uniform, with the pattern 
\begin{align} \label{eq:d_matrix_graph_2}
    \bm{t}^{\rm delay} = 
    \begin{pmatrix}
0 & 600 & 600 & 600 & 600 \\
600 & 0 &  630 & 630 & 630 \\
600 & 570 & 0 & 660 & 660 \\
600 & 570 & 540 & 0 & 690 \\
600 & 570 & 540 & 510 & 0  \\
    \end{pmatrix} \Delta t.
\end{align}
The general trends of entropy with symmetric or antisymmetric interactions shown in Fig.~\ref{fig:corr_entropy_graph_1}(b) and~\ref{fig:corr_entropy_graph_2}(b) are the same as those shown in Fig.~\ref{fig:delayed_entropy_5SMTJ}(b). Regarding the cross-correlation functions, Fig.~\ref{fig:corr_entropy_graph_1}(c-f) and~\ref{fig:corr_entropy_graph_2}(c-f) exhibit a consistent trend, where the antisymmetric interactions and non-zero delay times induce strong oscillations of correlation functions.

\clearpage
\bibliography{apssamp}%

\end{document}